# Designing an Optimal Scoop for Holloman High-Speed Test Track Water Braking Mechanism using Computational Fluid Dynamics

**Jose A. Terrazas**
University of Texas at El Paso
500 W University Ave, El Paso, TX 79968
jaterrazas2@miners.utep.edu

**Piyush Kumar**
University of Texas at El Paso
500 W University Ave, El Paso, TX 79968
pkumar2@utep.edu

**Arturo Rodriguez**
University of Texas at El Paso
500 W University Ave, El Paso, TX 79968
arodriguez123@miners.utep.edu

**Luis F. Rodriguez**
Clarkson University
8 Clarkson Ave, Potsdam, NY 13699
lrodrigu@clarkson.edu

**Richard O. Adansi**
University of Texas at El Paso
500 W University Ave, El Paso, TX 79968
roadansi@miners.utep.edu

**Vinod Kumar**[1]
Texas A&M University - Kingsville
700 University Blvd, Kingsville, TX 78363
vinod.kumar@tamuk.edu

[1] Corresponding author: vinod.kumar@tamuk.edu

## ABSTRACT

*Specializing in high-speed testing, Holloman High-Speed Test Track (HHSTT) uses "water braking" to stop vehicles on the test track. This method takes advantage of the higher density of water, compared to air, to increase braking capability through momentum exchange by increasing the water content in that section at the end of the track. By studying water braking using computational fluid dynamics (CFD), the forces acting on tracked vehicles can be approximated and prepared before actual testing through numerical simulations. In this study, emphasis will be placed on the brake component of the tracked sled, which is responsible for interacting with water to brake. By discretizing a volume space around our brake, we accelerate the water and air to simulate the brake coupling relatively. The multiphase flow model uses the governing equations of the gas and liquid phases with the finite volume method to perform 3D simulations. By adjusting the air and water inlet velocity, it is possible to simulate HHSTT sled tests at various operating speeds.*

## INTRODUCTION

The Holloman High-Speed Test Range (HHSTT) is operated by the 846th Test Squadron (TS) at Holloman Air Force Base (HAFB) near Alamogordo, New Mexico. It is the world's premier facility for testing full-scale, low-altitude, high-speed flight vehicles and conducting cutting-edge test and evaluation (T&E) capabilities to support and develop critical weapons systems technology [1,2]. HHSTT performs a wide range of tests to evaluate the performance of various aerospace systems, including bomb and fuse functions, impact resistance, weapons lethality, ejection seat performance, guidance system accuracy, material erosion, ablation, the impact of rain on aerospace materials durability, aerodynamic properties of aerospace system designs, and others [3–5]. It has been used to perform hypersonic (beyond Mach 8) impact tests [4,6–10]. Using rocket-powered sleds running on a 9.63-mile steel track in the past. Steel shoes or skids support the rocket sleds on the rails [11,12].

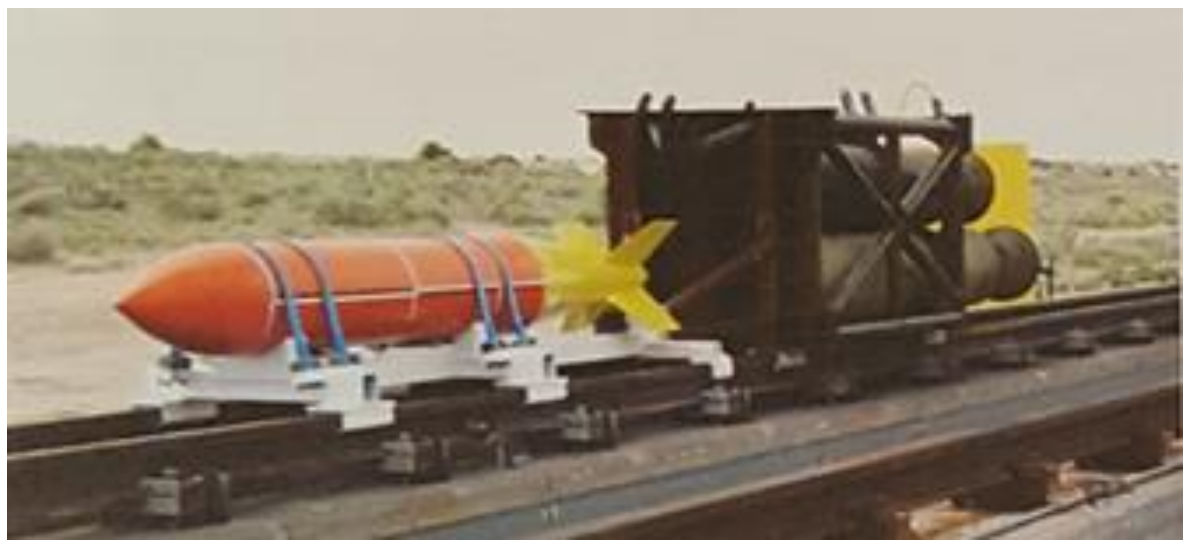

Figure 1: Test article and pusher of the sled design

Operational flight speed conditions are achieved using rocket-powered sleds mounted on rails. Sleds consist of a nose component, where the tested item sits. The braking system and push component accelerate the nose to the desired speed. Because many tests require recovery of the test item for post-test analysis, braking mechanisms with sufficient capacity to stop the sled are needed. Braking methods to prevent or recover sleds include cambering the track, water braking, and ailerons such as air brakes, hooks, and cables [13]. Various methods of implementing water braking depend on the test speed. Depending on the speed requirements, the sled will be propelled by either a dual rail or a monorail, which affects how water braking is performed. For dual rail push sleds [4,14,15], water is deployed at increasing depths in a channel between the rails. Water depth is controlled using frangible dams inserted into the channel at planned locations and distances. The sled's brake paddle encounters and deflects water in the channel, providing a drag force that slows the sled to a stop.

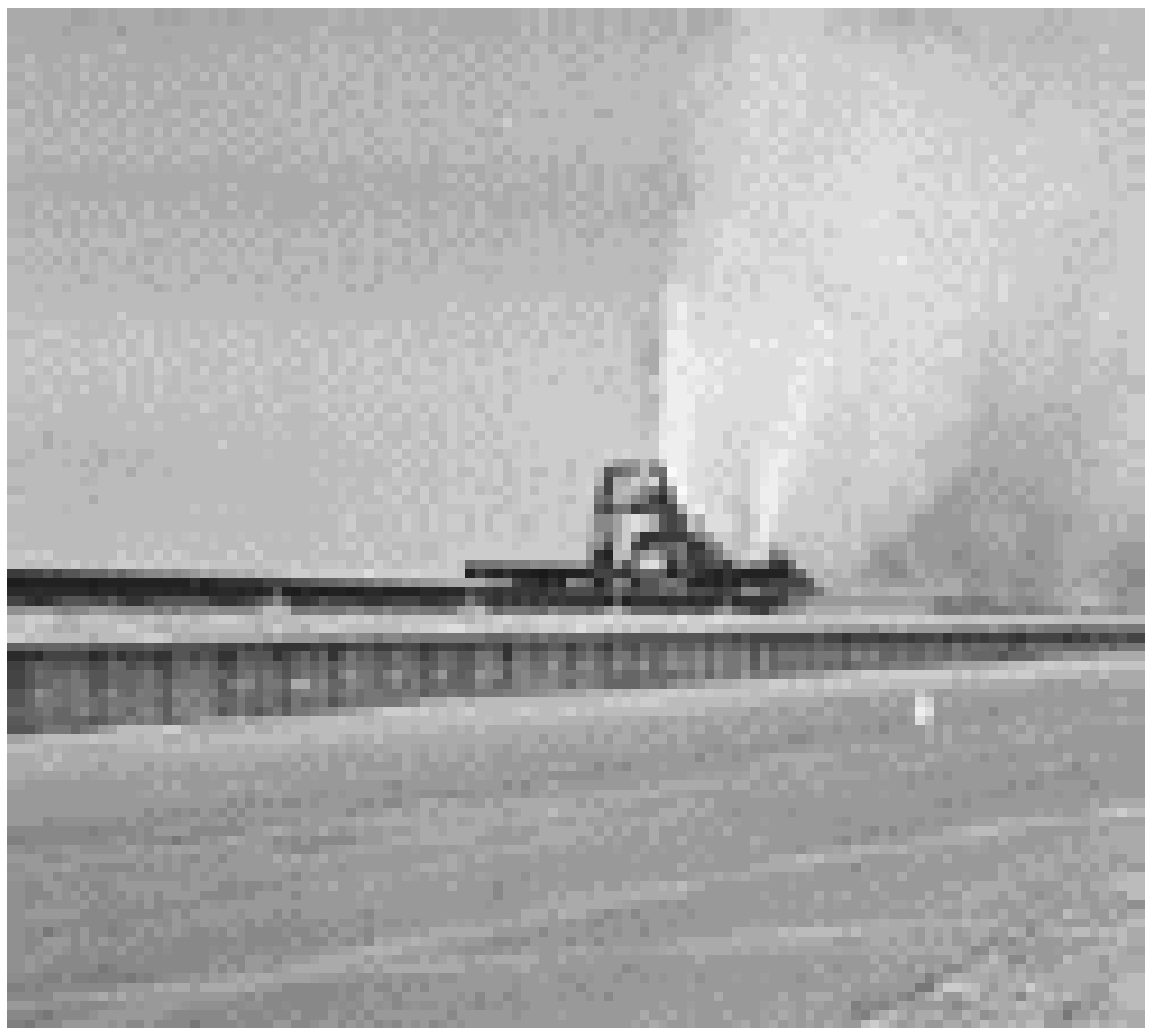

Figure 2: Water braking at Holloman High-Speed Test Track

In recent decades, CFD simulation tools have developed and become powerful tools for analyzing and understanding fluid dynamics phenomena. Due to the further development of CFD tools, Air Force system designs, such as the high-speed sleds at HHSTT, can be improved[16,17]. The credibility of CFD simulations can only be established through a rigorous verification and validation process [18–21]. Using CFD, engineers, and scientists can analyze and study fluids quickly, cost-effectively, non-intrusively, and in a feasible manner compared to running precise tests and experiments. To this end, CFD modeling is an excellent tool for the US Air Force (USAF) at HHSTT because they work with push sleds to test vehicles under high-speed conditions. CFD allows for better engineering, construction, and predictability of the braking files of the tested vehicles. Improving the design of push sleds allows for better braking performance and reusability, decreasing the cost of track testing and increasing the speeds at which vehicles can be tested.

Over several decades, the HHSTT has conducted several tests of high-speed sleds with water-braking mechanisms and collected data on predicted profiles with varying degrees of success, confidence, and fidelity. We followed established guidelines for assessing the credibility of CFD simulation and modeling results [18,22–24]. The geometry, computational domain, and boundary conditions for the three-dimensional numerical simulations are presented in Problem Definition. The following information delves into the two- and three-dimensional results, analysis, and discussion of dual-rail water braking, followed by a conclusion and a final comparative study of all designs.

**PROBLEM DEFINITION**

Improving the prediction capabilities of water braking phenomena can lead to radical changes in sled designs, improve predictions of the velocity-time test profile of rocket sleds, provide greater confidence in braking mechanisms, and reduce risk in critical infrastructure recovery [25,26]. Understanding water behavior with the sled is essential to predict how water might damage the sled, which impacts its recoverability and can determine the success of a mission and the amount of drag it will experience from air and water. Traditionally, sled design for water braking test missions has been guided by empirical/manual calculations to estimate the forces on various components. The calculations involve various approximations to determine the force balance law and predict the acceleration/deceleration profile. CFD results of different geometric configurations for the sled and modeling parameters will be presented. The main objective of CFD investigations is to improve the accuracy of the predicted profile, which often depends on the complexity of the design and the operating conditions.

In summary, five different geometries of the sled water braking mechanism, the scoop, are simulated in a 3-dimensional space. The analysis of scoop simulations began with 2D cases, where conditions and solvers were studied before moving on to 3D cases. The objective was to become familiar with scoop simulation and ANSYS CFD Fluent software before performing 3D modeling, which is much more time-consuming and computationally expensive. The article is divided into 2D and 3D models. Each division has its geometric sections, discretization statistics, and results analysis unique to each scoop design.

For the three-dimensional setup, the concept and boundary conditions for all models can be understood from Figures 3 and 4 [26,27]. In Figure 3, we have boundary conditions with water and air inlets from the left, pressure outlets from the right, and walls for the top, bottom, and scoop. The difference between all the models is geometric changes to the scoop; all other conditions were kept the same. The turbulence model used for all was the k-epsilon 2 equation RANS model, with a SIMPLE numerical scheme. Data analyzed from each simulation were pressure, volume fraction, velocity, vorticity, turbulent kinetic energy, turbulent intensity, and volumetric flow rate. Consistency in mesh quality, domain, turbulence model, numerical scheme, and transient specifications was maintained to allow valid comparisons between geometries.

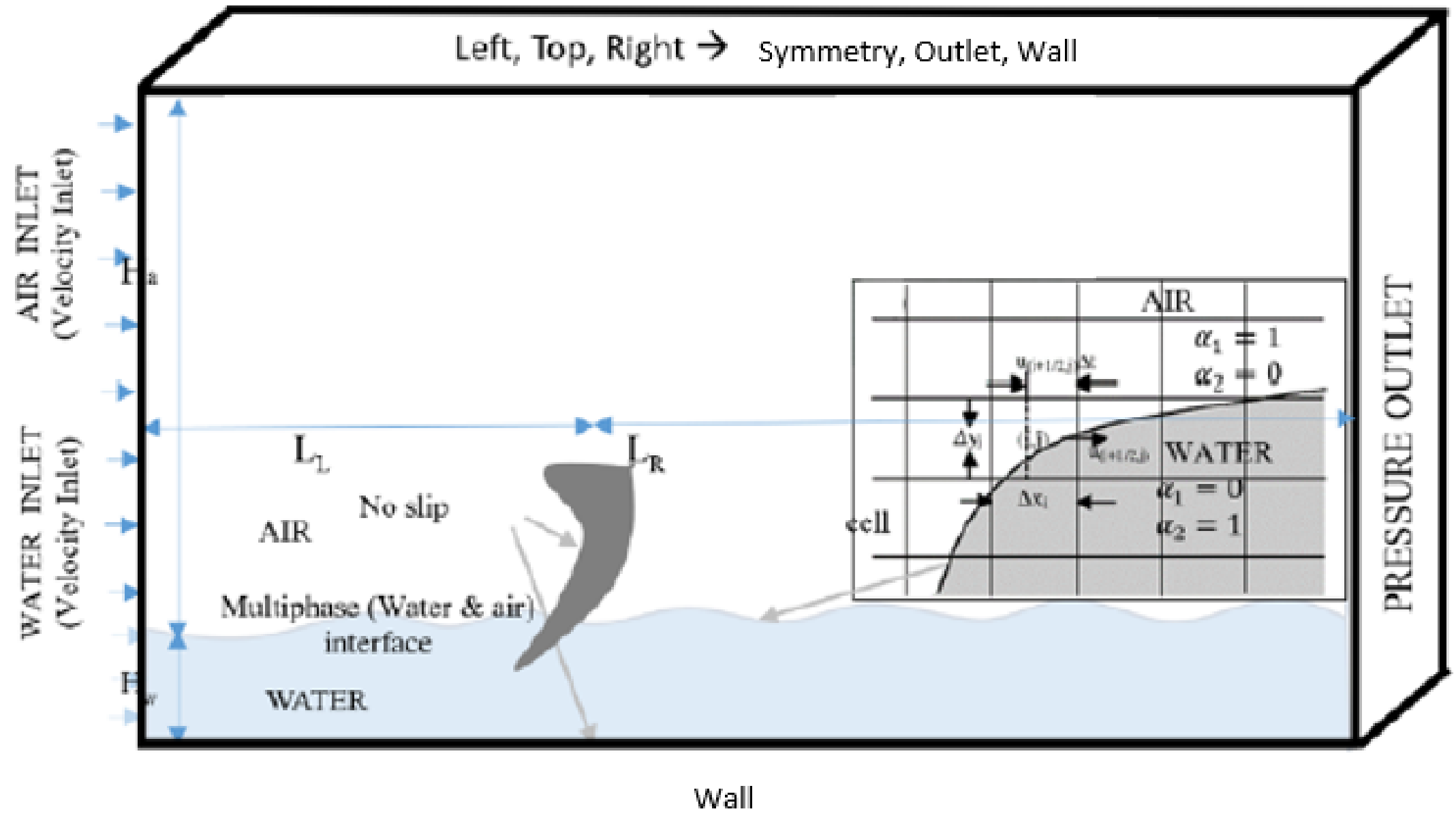

Figure 3: 3-D boundary conditions and set-up

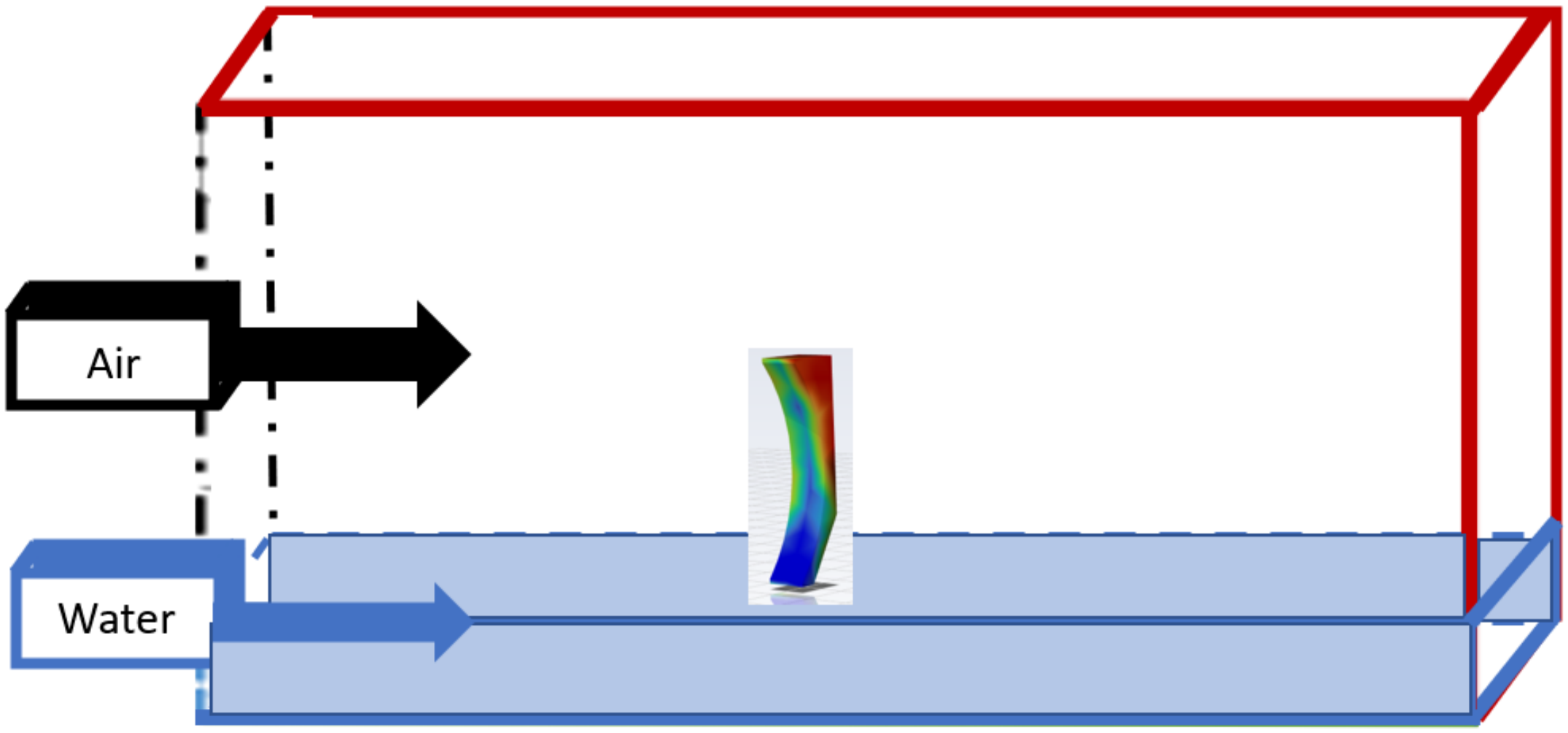

Figure 4: 3-D boundary conditions and set-up

**METHODOLOGY | 2-D MODELING**

The initial 2-D scoop design is inspired by channeling water flow through the scoop while providing a 60-degree divergent geometry behind the bottom to reduce the lift force caused by water flowing under the scoop (see Figure 5).

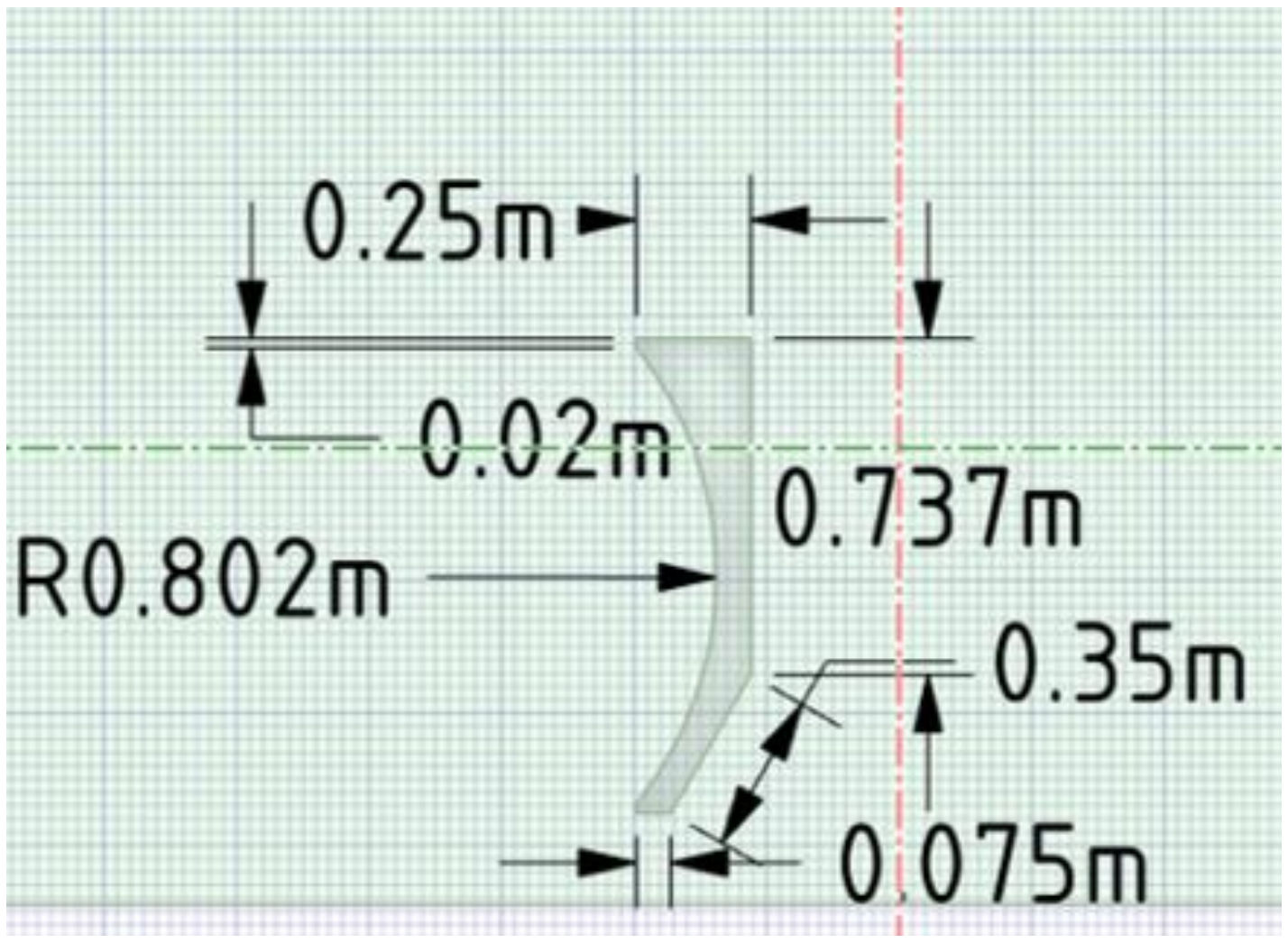

Figure 5: 2-D Design.6 Scoop geometric dimension

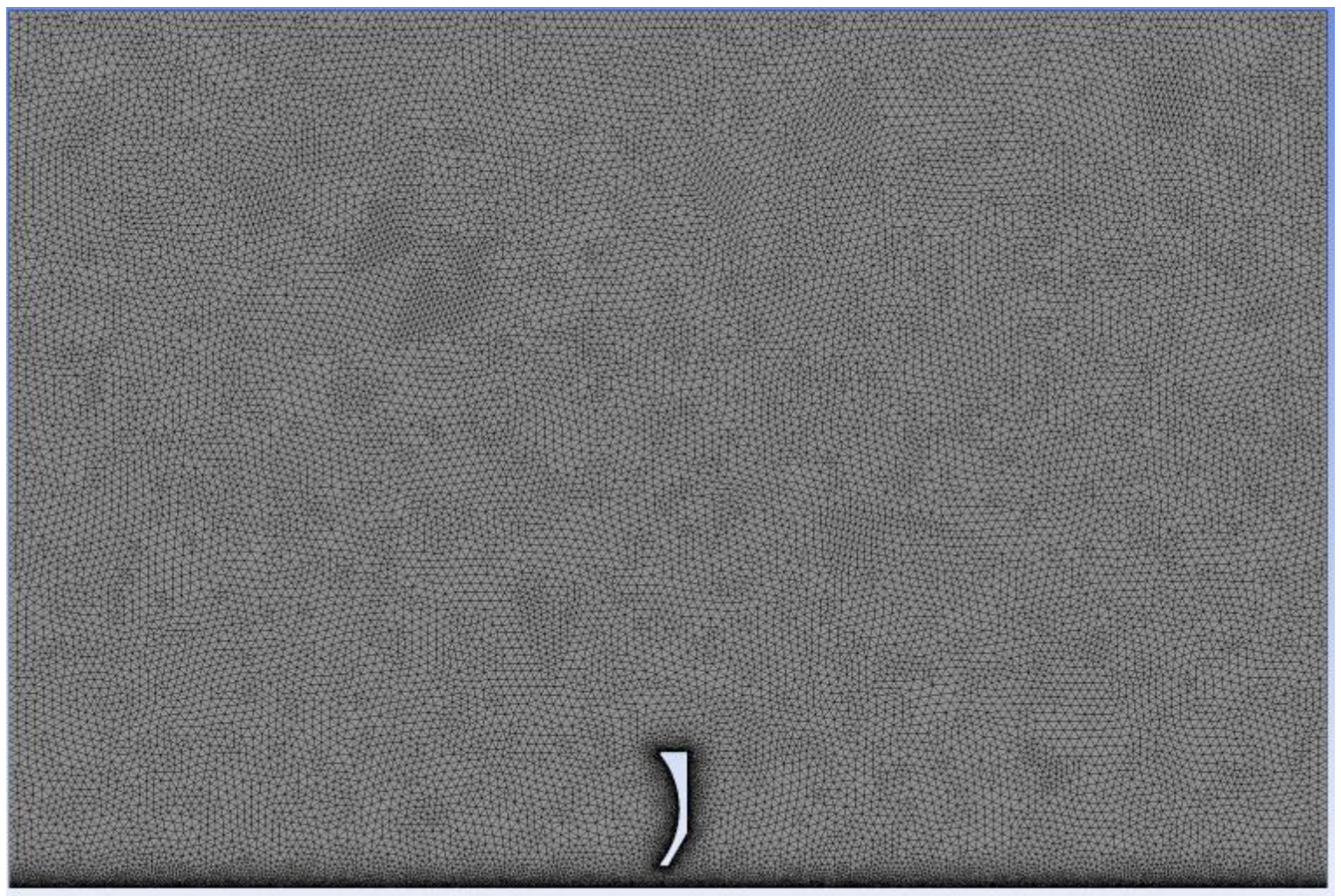

Figure 6: 2-D Design.6 mesh

The discretization of the 2-D model was focused mainly on greater fidelity around the scoop and water channel, as most of the force will be coming from the water (see Figure 6). (2-D models are geometrically 8 meters by 12 meters.) Element quality was very high, with an average of 0.96229, ensuring quality results. However, an issue with element quality, specifically with complex geometries and unstructured meshes, is the minimum quality of an element. Due to some small geometric details, obtaining an acceptable minimum-quality element can take time and effort. Even a few poor elements can be detrimental to a simulation and lead to numerical instability that keeps it from converging and reaching a steady state.

**RESULTS AND DISCUSSION | 2-D MODELING**

Qualitative results align with what is expected from the initial airflow on the scoop seen in Figure 7. Figure 7 shows the contour of the Mach Number and density of the air when the scoop experiences contact with the water early in the simulation. Later in the simulation, the drag force captured a force of 8.79E6 N at an inlet velocity of 300 m/s after water contact.

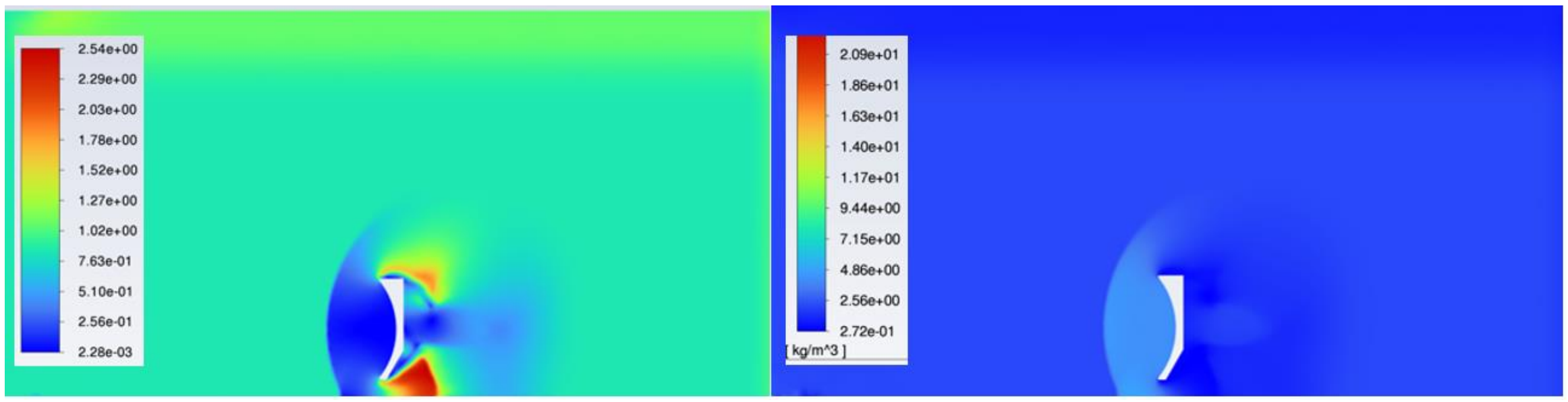

Figure 7: Mach number (left) and density of air (right) of 2-D Design.6 at 300 m/s

**VERIFICATION | 2-D MODELING**

The 2D model in Figure 6 illustrates the control volume (domain) and fluid flow. It was run with a 300 m/s velocity inlet for air and water. As previously stated, the purpose of the 2D model was to establish initial and boundary conditions. Using Figure 7, calculations by hand can be estimated at 300 m/s and compared to software simulations to validate. The results are shown below.

Force (by hand) = 9.0e6 N/m     Force (2D model) = 8.7e6 N/m

Although 2-D modeling is essential for establishing initial conditions, boundary conditions, and competence in ANSYS Fluent, it has little utility from an analysis standpoint for a real-world condition. Being 2-D means it lacks depth in which water and air can flow, changing the calculus of scoop geometric studies. A 2-D model may be viewed as a geometry with infinite depth.

**METHODOLOGY | 3-D MODELING**

The first step after 2-D modeling was to determine different geometric changes using a 3-D version of the 2-D model [27]. Five geometric variations, including the 3-D adaptation of the 2-D scoop, were analyzed in the 2-D modeling section. All geometries have some modifications from each other, starting with Design.6. Design.6 is the first to be modeled in 3D for this study; it is a direct model from the 2D scoop (see Figure 5) but with a 0.2-meter depth.

From Design.6 to Design.7, for example, "guards" were added to the sides of the front of the scoop to channel water in the scoop. In Figure 8, all five different variations are illustrated. Green demonstrates the addition of the previous design, while red is the removal of a particular geometric distinction. Figure 9 shows the front and side views without pointing out geometric differences like in Figure 8. Figure 9 allows a clearer individual picture of the different scoop designs.

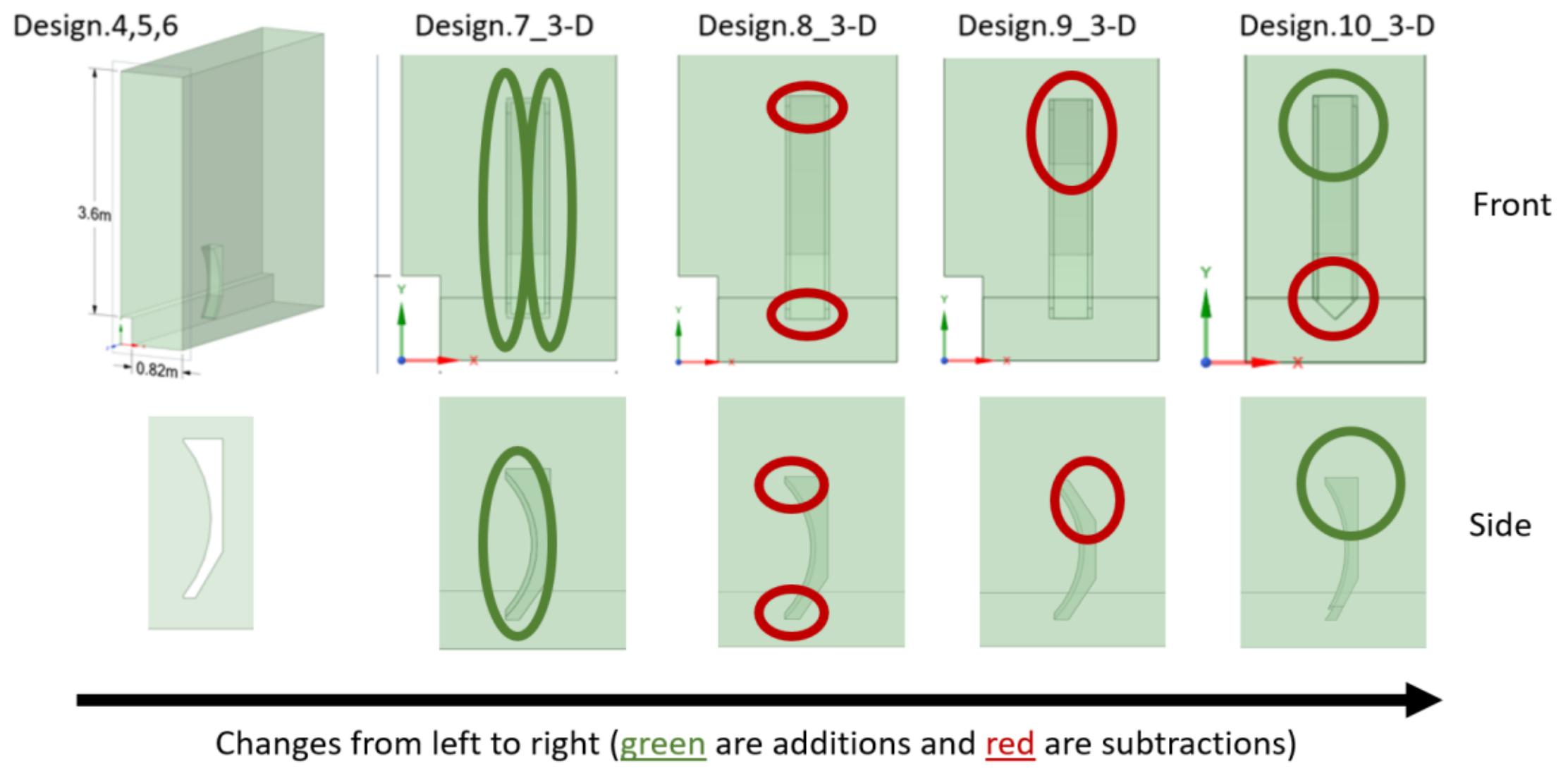

Figure 8: Variations between the different 3-D scoop designs

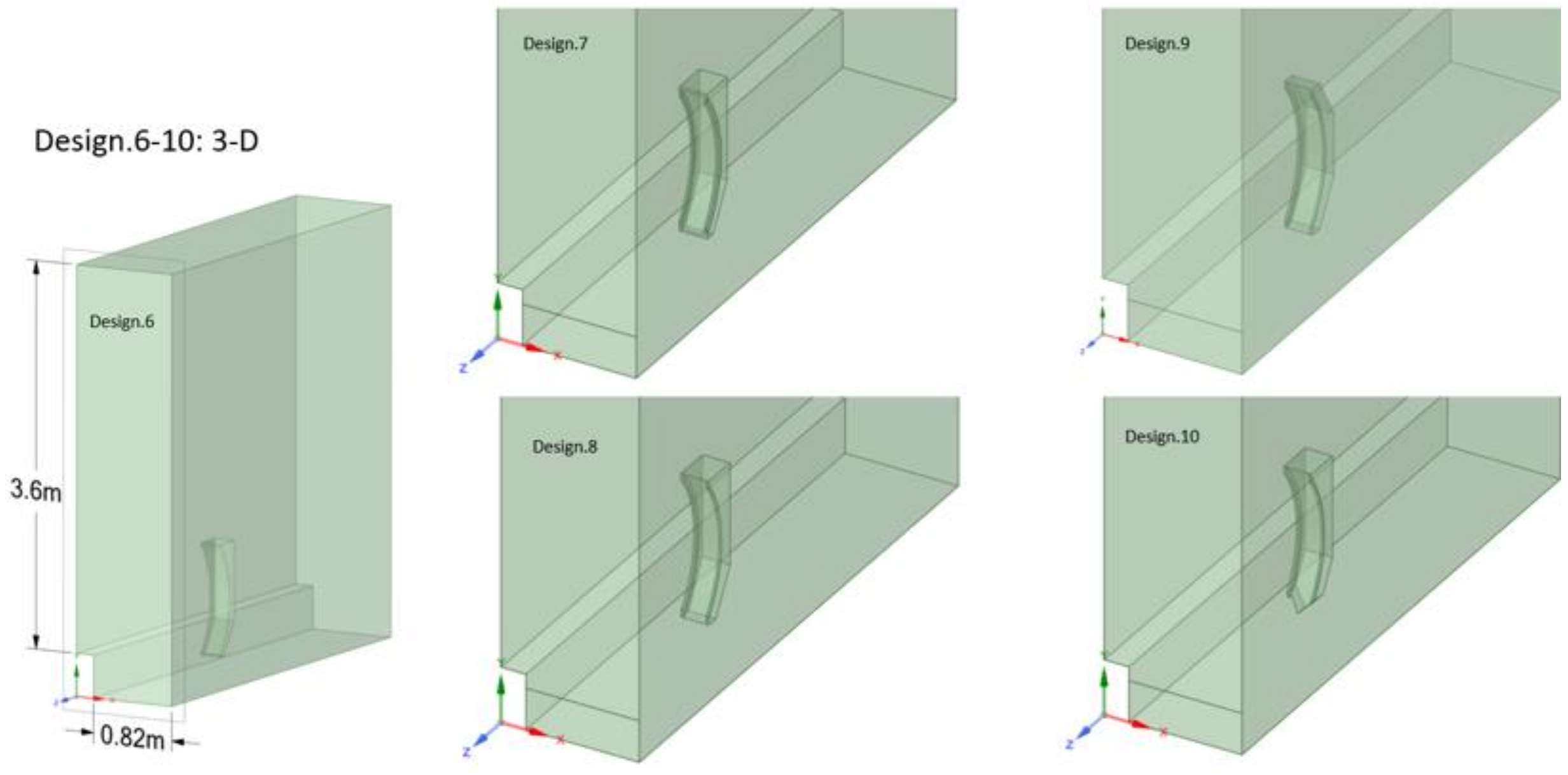

Figure 9: Isometric view of the different scoop designs

To that end, the same type of cell topology, tetrahedrons, was used, as were efforts to obtain a similar mesh quality. Geometric differences meant that having the same mesh statistics was impossible, but they were all within 1% of each other regarding the number of elements. The specified element size was the same for all. All five scoop designs, starting with Design.6, were simulated at 100 m/s. Due to computational resource limitations, any faster flow was complex to converge with mesh limitations. Only Design.9 was modeled successfully to an extent of 300 m/s. The following sections of 3-D modeling go into the details of the results for each design. Qualitative data in the following contours are extracted from a plane going across the domain on the Z-axis.

## RESULTS AND DISCUSSION | 3-D MODELING

### Design.6 at 100 m/s

Figure 10 shows the volumetric flow rate of water in front, isometric, and side views. It illustrates how the water engages explicitly with the scoop and how the water flows. Critical to this analysis is the inspiration for geometric changes to attempt to channel water flow in a more productive capacity for our purposes, which is to increase drag.

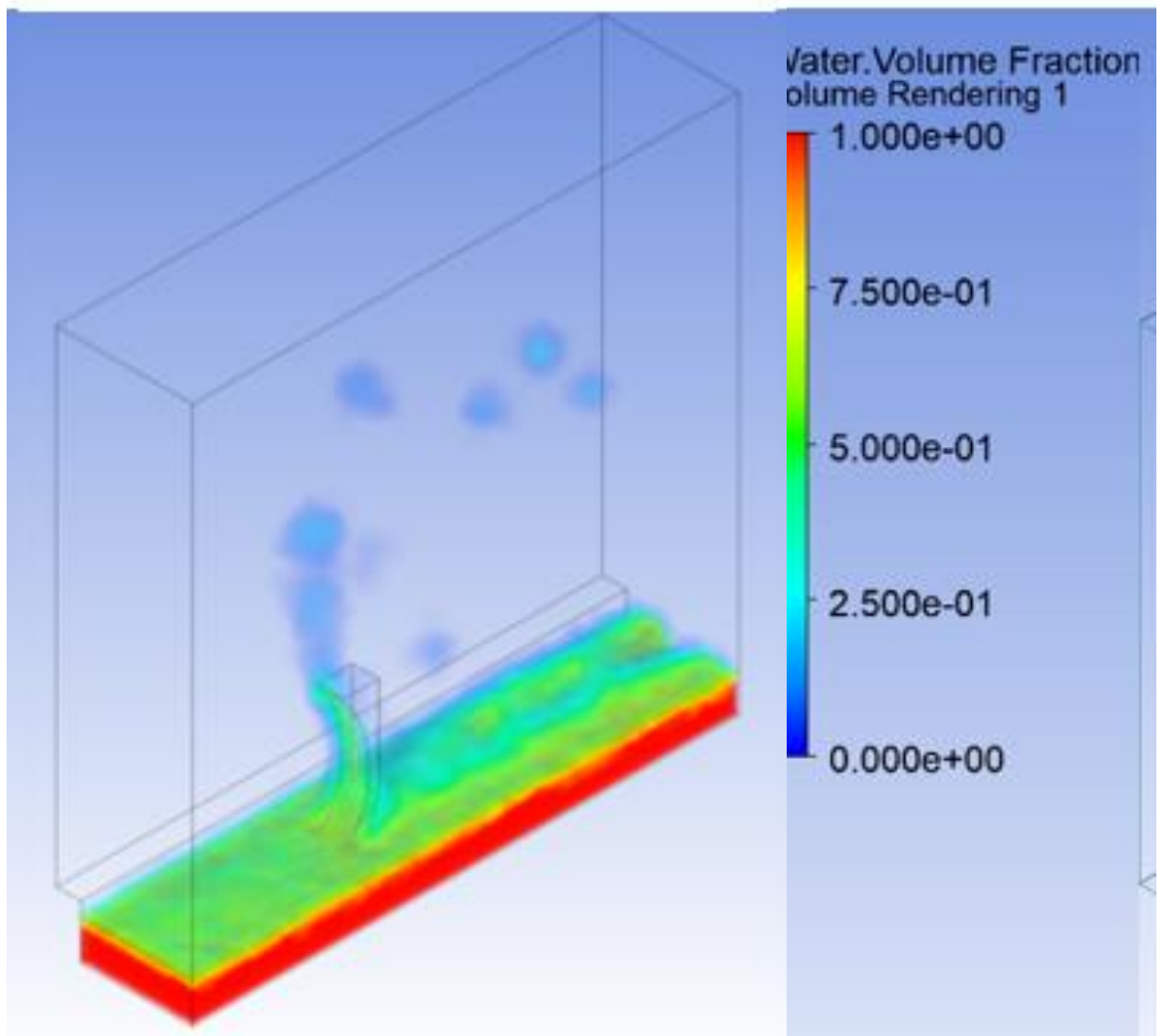

Figure 10: Design.6 water volumetric flow rate

In Figure 11, we have a quad chart of velocity contours; careful interpretation of 3-D results requires noticing the orientation of the X, Y, and Z axes. The Z-axis follows the fluid flow direction inversely, but adjustments were made to make it favorable in the legend. The X-axis is the "depth" in the domain, and the Y-axis is the height.

From the velocity quad chart, starting with the top left, we have magnitude, followed by velocity in the Z-axis on the top right, velocity in the X-axis on the bottom left, and velocity in the Y-axis on the bottom correct. Also, the velocity contours and the pressure are through a plane across the domain, which provides a picture of what is happening around the middle of the domain. As expected, most of the velocity flow is on the Z-axis, but there is flow in the X and Y directions, as well, that impacts overall flow. Pressure is also highly concentrated at the bottom of the scoop, where water first engages it.

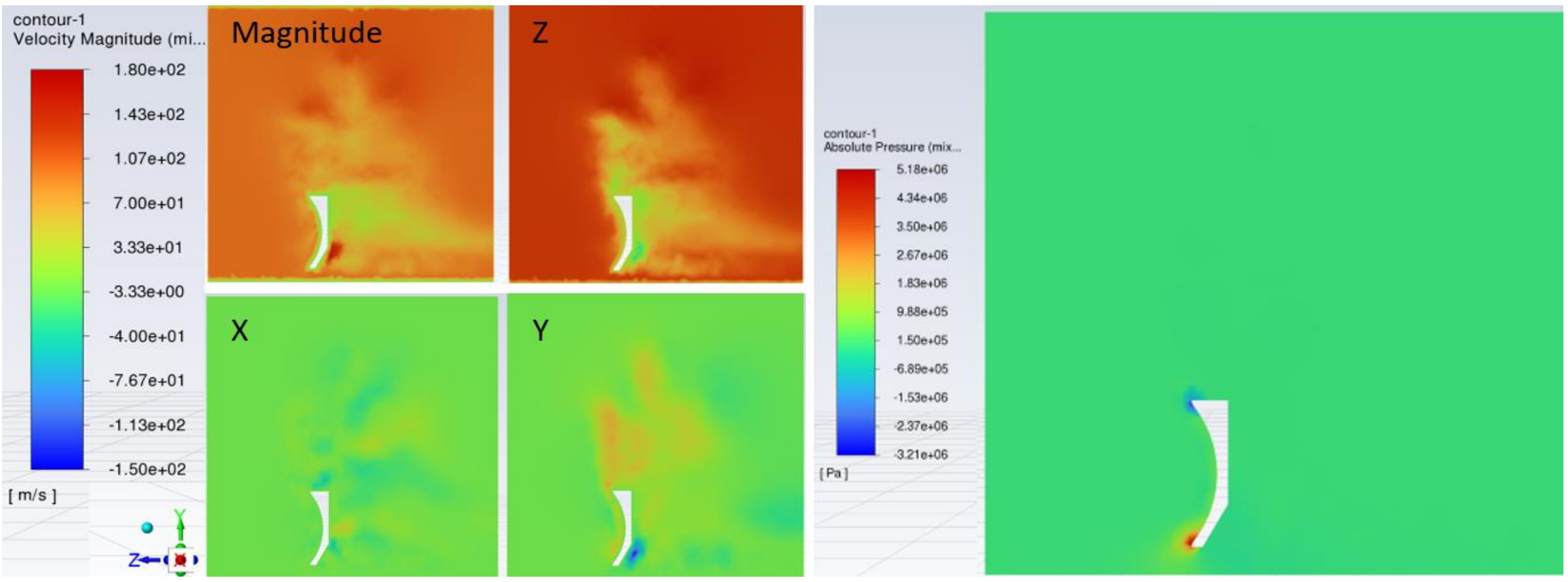

Figure 11: Design.6 velocity contours (left) and pressure contours (right)

Turbulent kinetic energy, in conjunction with turbulent intensity and vorticity, is essential in determining the amount of turbulence generated and, hence, the drag on the scoop. By viewing the contours, we can distinguish where the highest concentration is and make geometric adjustments to improve drag on the scoop.

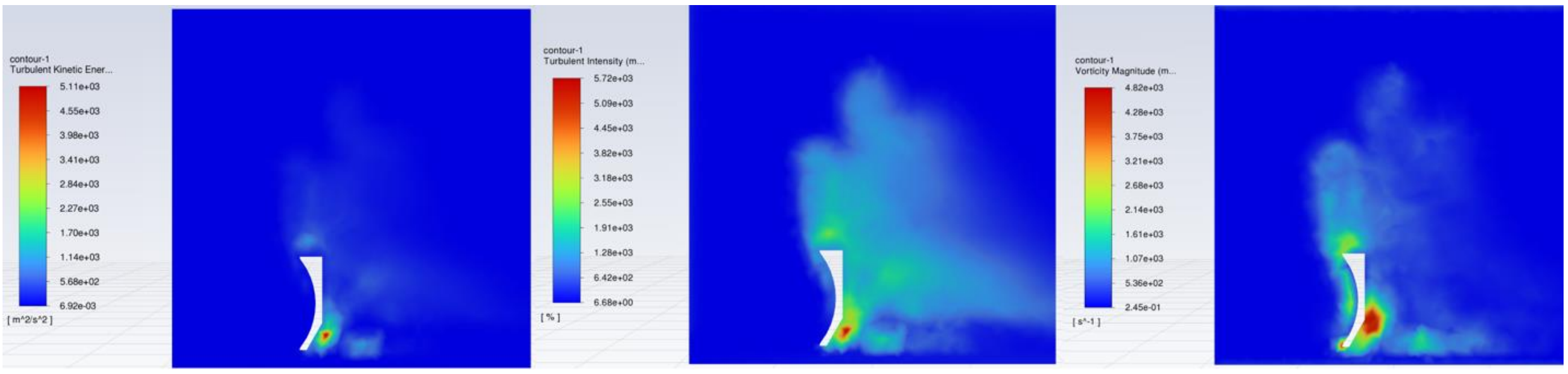

Figure 12: Design.6 Turbulent Kinetic Energy (KE) contours (left), Turbulent Intensity contour (center), and Vorticity magnitude contour (right)

The force on our initial 3-D model is 139 kN, as seen in Plot 3.1. Also, in Plot 3.1, we can see the quantification of the pressure observed in Figure 11, and by cross-examining them, we can distinguish the regions of high pressure and low pressure in the geometry.

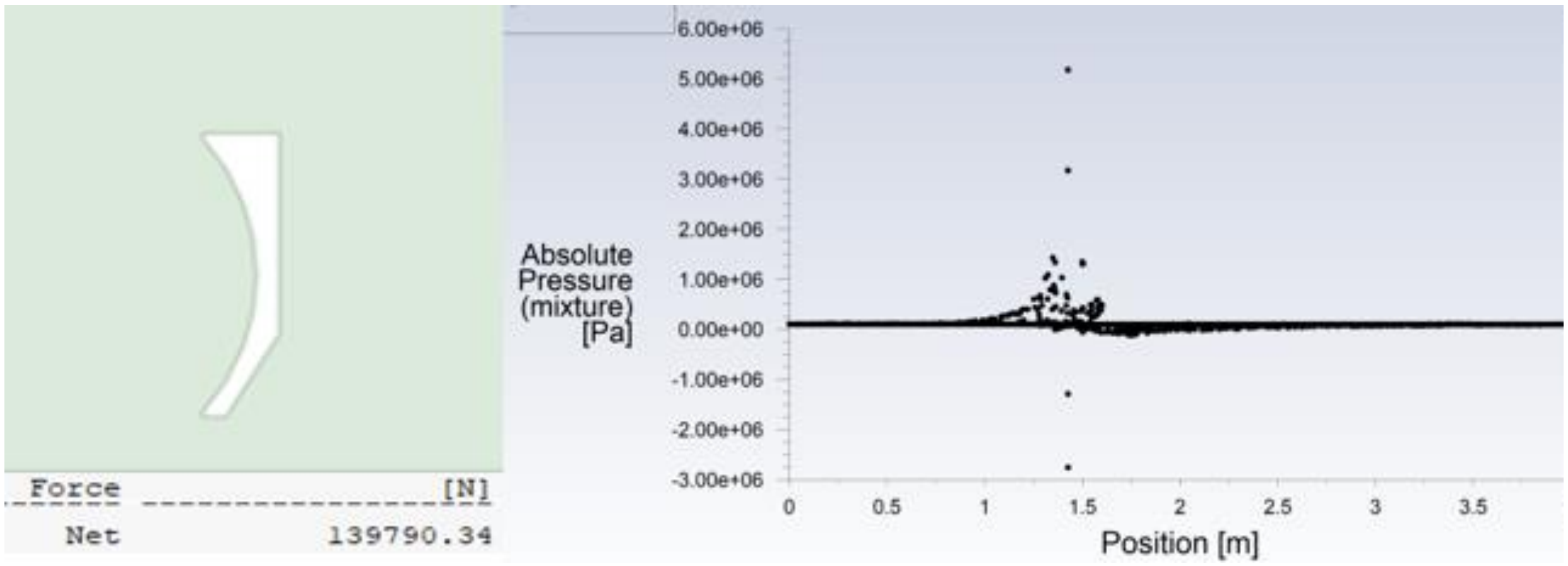

Plot 1: Design.6 side view, drag force (left), and absolute pressure plot (right)

Like comparing a pressure plot with its contour, an analysis of vorticity, velocity, turbulent kinetic energy, and turbulent intensity plots in Plot 1 can be made with Figures 11 and 12. Knowing that the scoop is at the 1.425-meter mark and goes to the 1.675-meter mark, we can conclude that the peaks in turbulence are occurring past the arc of the scoop and that a lot of the water flow is being lost to the sides of the scoop. With this information, we can proceed to a new design, Design.7.

**Design.7 at 100 m/s**

In Design.7, we can take lessons learned from Design.6 and apply them to future designs, understanding that contours should be standardized into Design.7 and onwards as contours become part of the analysis. According to the analysis conclusion of Design.6, “guard rails” on the front arc of the scoop are integrated to hopefully guide the water flow through the scoop and generate greater drag.

In Figure 3.18, the green ovals show exactly how these guard rails will be integrated and how thick they will be. The only changes from Design.6 to Design.7 are 0.3-meter guard rails on both sides of the front arc of the scoop, which retain a small flat surface at the top and bottom.

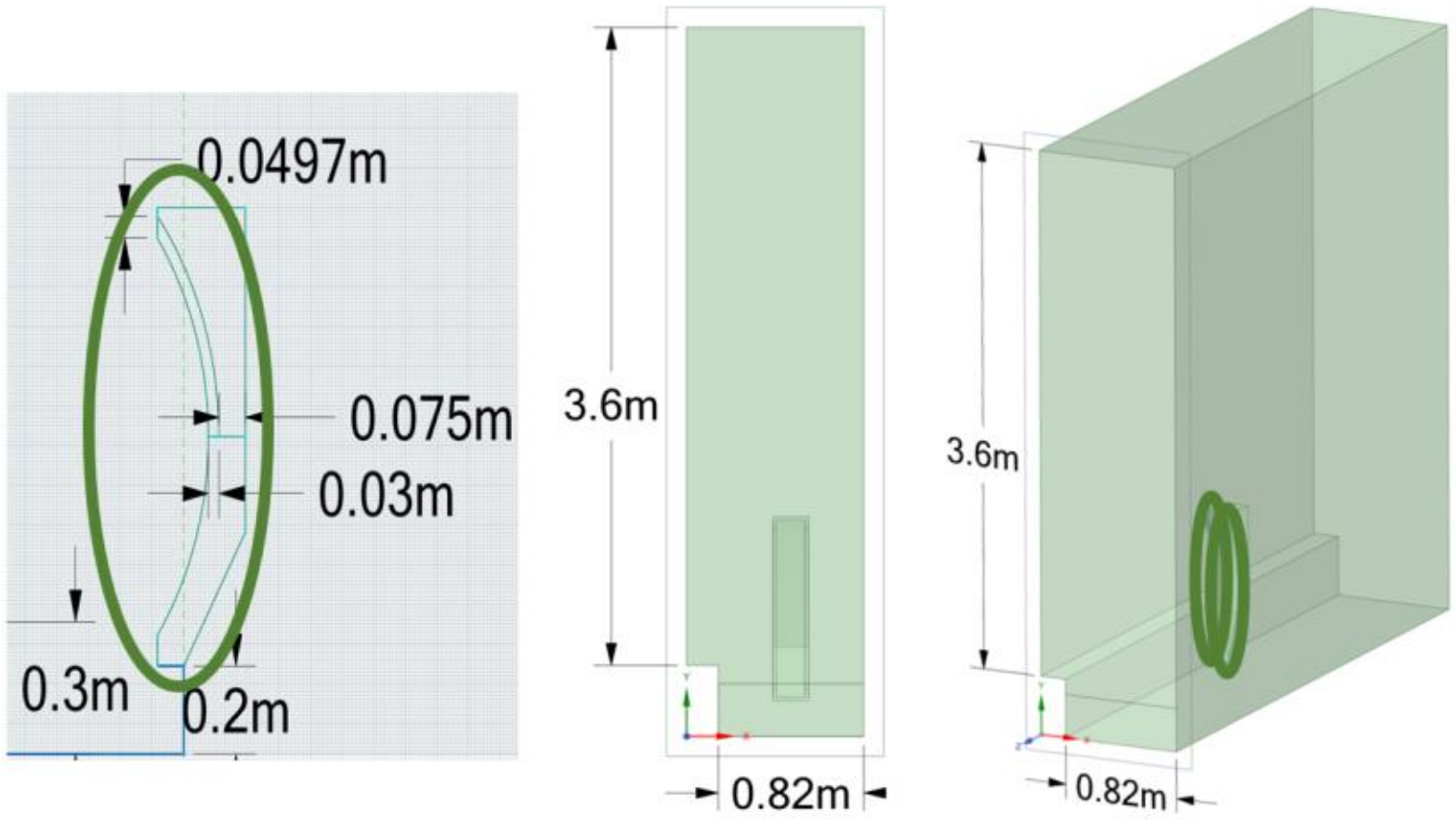

Figure 13: Design.7 geometry changes and views from changes made from Design.6

Learning from Design.6 for Design.7, data was explicitly designed to integrate contour and plot. As seen in Figure 3.19, the pressure contour and plot are now together and at the exact dimensions (the plot encompassing the domain range) to better view parameter behavior on the domain plane. As in Design.6, there are pressure spikes at the front of the scoop where water initially contacts the scoop. Low-pressure areas are on top of the scoop but are now also more defined behind the scoop. Comparing the pressure plots of Design.6 and Design.7, it is already evident that the force will be much larger in Design.7. In Design.7, the peak pressure registered is more than 30% higher than in Design.6.

For contours to follow, in addition to being coupled with their respective plots, the legend is adjusted to be the same going forward into Design.8, Design. 9, and Design. 10. This will facilitate qualitative and quantitative comparisons between the results of each design, making it easier to identify improvements.

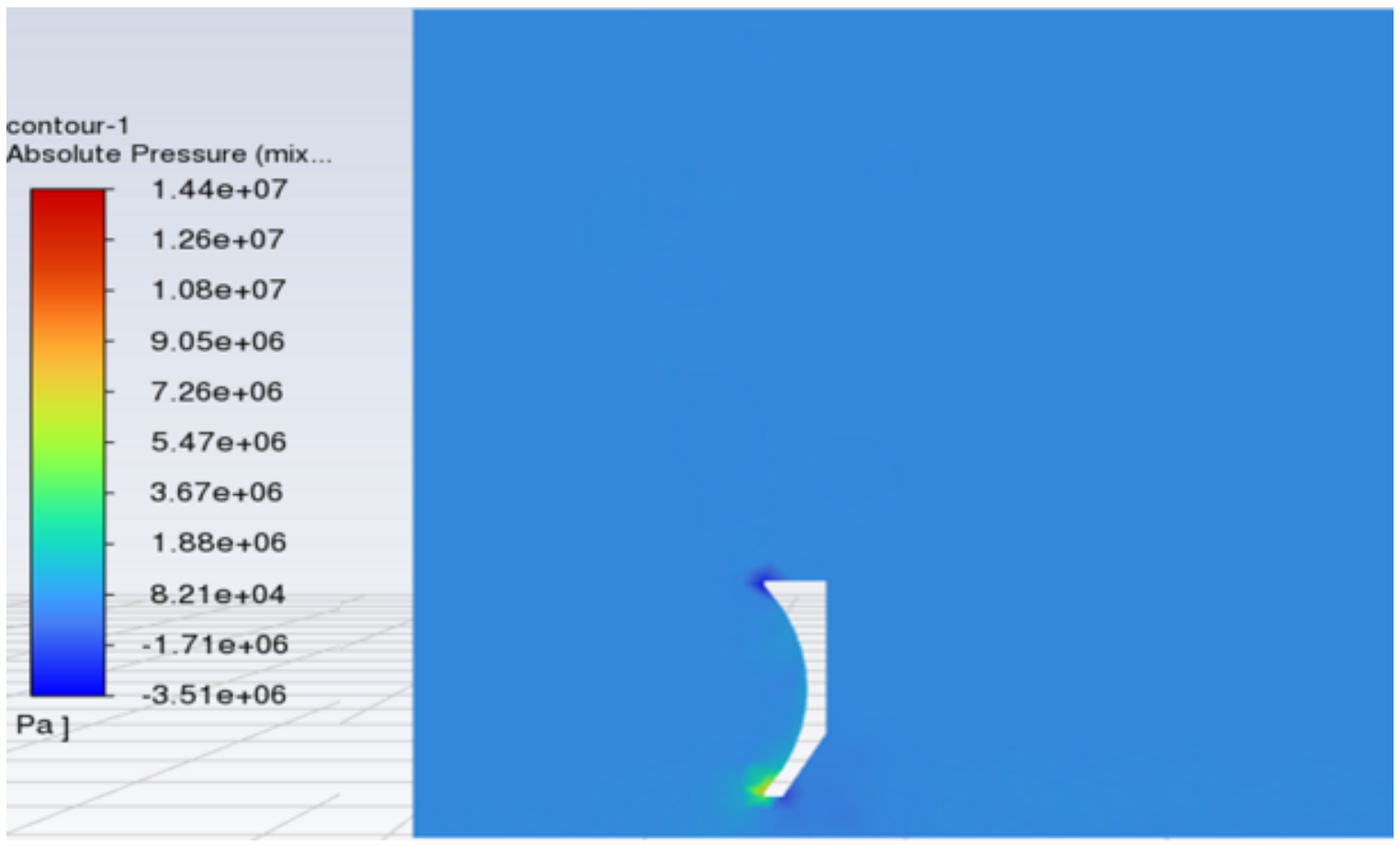

Figure 14: Design.7 absolute pressure contour (top) and plot (bottom)

For velocity magnitude, looking at Figure 15, there are higher velocities than in Design.6. It is important to remember, however, that velocity points are mixed with water and air. A better picture will be formed by analyzing velocity and vorticity fields in the X, Y, and Z axes, not just magnitude. Magnitude velocity and vorticity, however, indicate what is expected and how the flow has changed since the guard rail addition. For vorticity (Figure 3.21), the flow control seems to have concentrated it around the scoop instead of further along the domain.

Turbulent kinetic energy (TKE) and turbulent intensity (TI) (Figure 15) are difficult to compare since contours and plots are not the same. Still, they are higher in Design.6. This could be due to the mixture of air and water in the data. Turbulent kinetic energy (TKE) and turbulent intensity (TI) will be most important in the analysis behind the scoop as the turbulent flow contacts scoop attachments such as the pusher sled. The priority parameters for the scoop will be pressure and force.

In Figure 16, we can observe the volumetric flow rate of water and force over time. The red line in the bottom right force plot indicates the moment water contacts the scoop. From the water contour, there is much higher water concentration on the arc of the scoop than in Design.6, which can only be attributed to the guard rails. A significant increase in drag force is also demonstrated at 195 kN, or about a 40% increase.

For Design.7, many more data plots and contours were created to extract more detail on the flow. Details that were not extracted in Design.6 include vorticity plots in all axes (Plot 1), velocity plots in all axes (Plot 1), and finally, a force plot as it relates to the height of the scoop (Plot 1). This data will be used to make more detailed comparisons with the following designs.

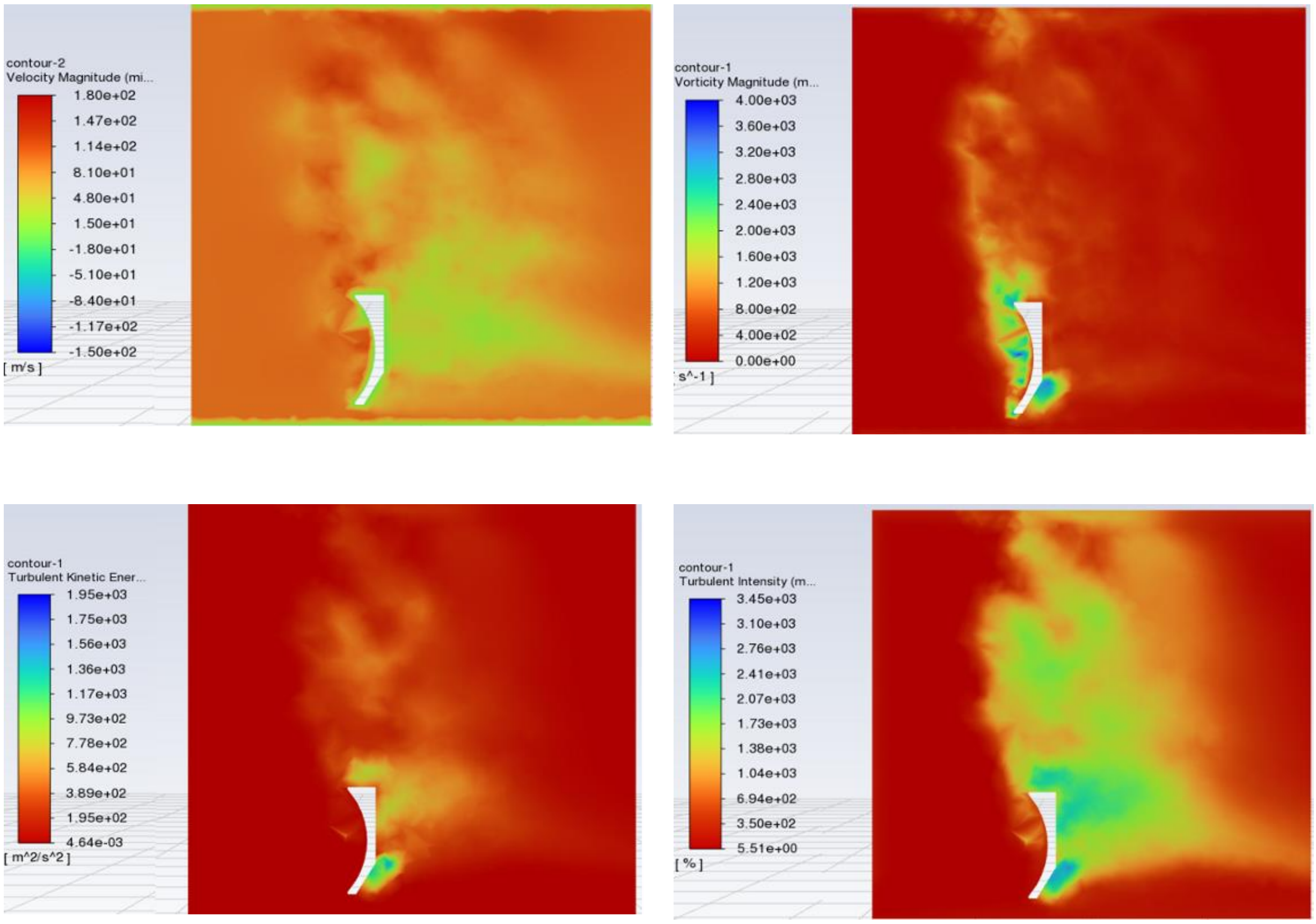

Figure 15: Design.7 (a) velocity magnitude contour (b) vorticity magnitude contour (c) turbulent kinetic energy contour (d) turbulent intensity contour

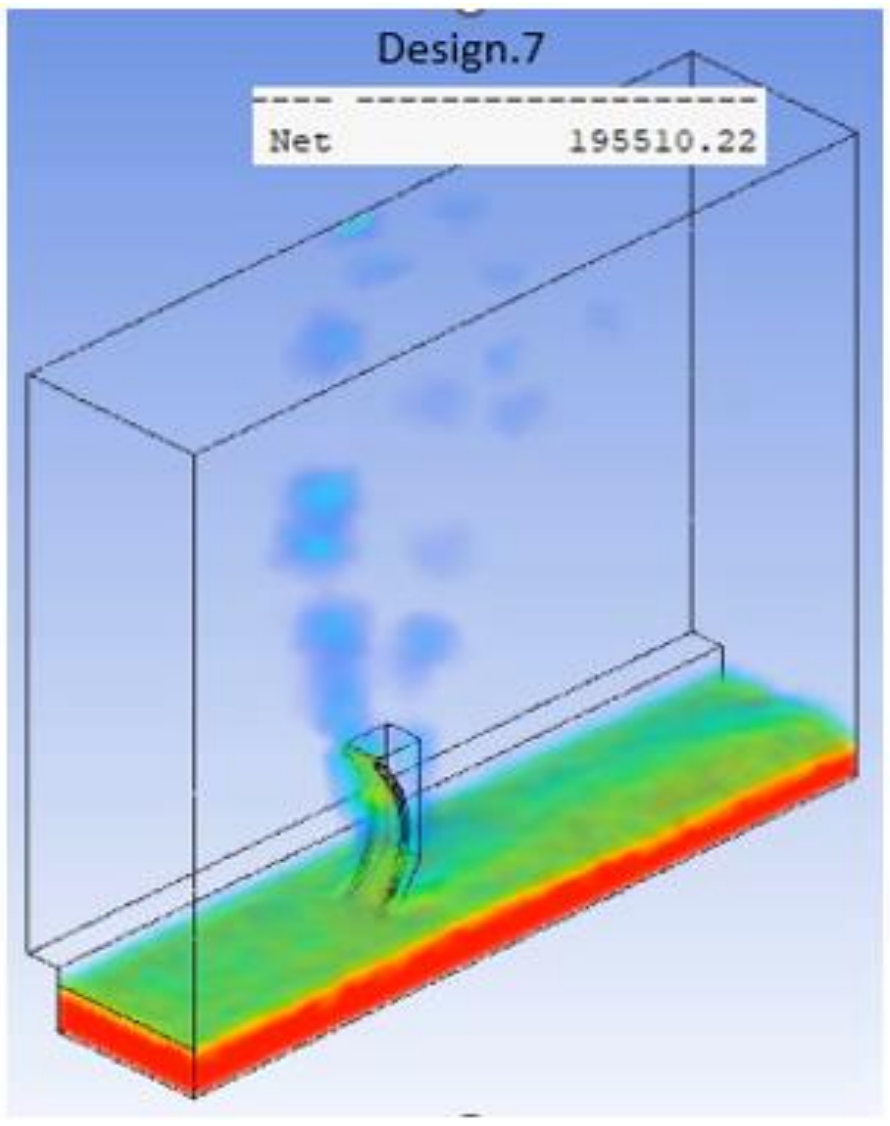

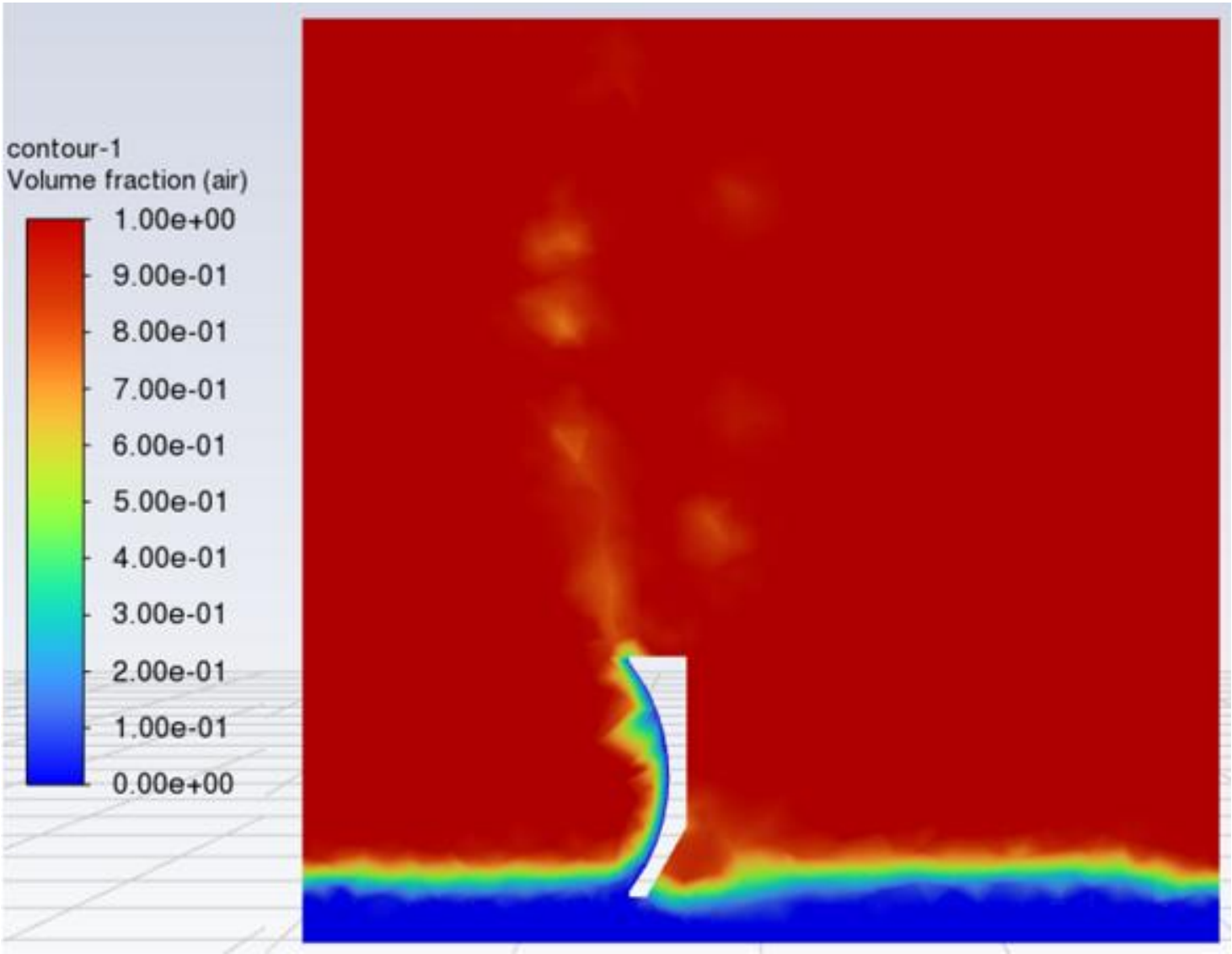

Figure 16: Design.7 (a) water volumetric flow rate (isometric view) (b) volume fraction of air in a plane through the middle of the domain

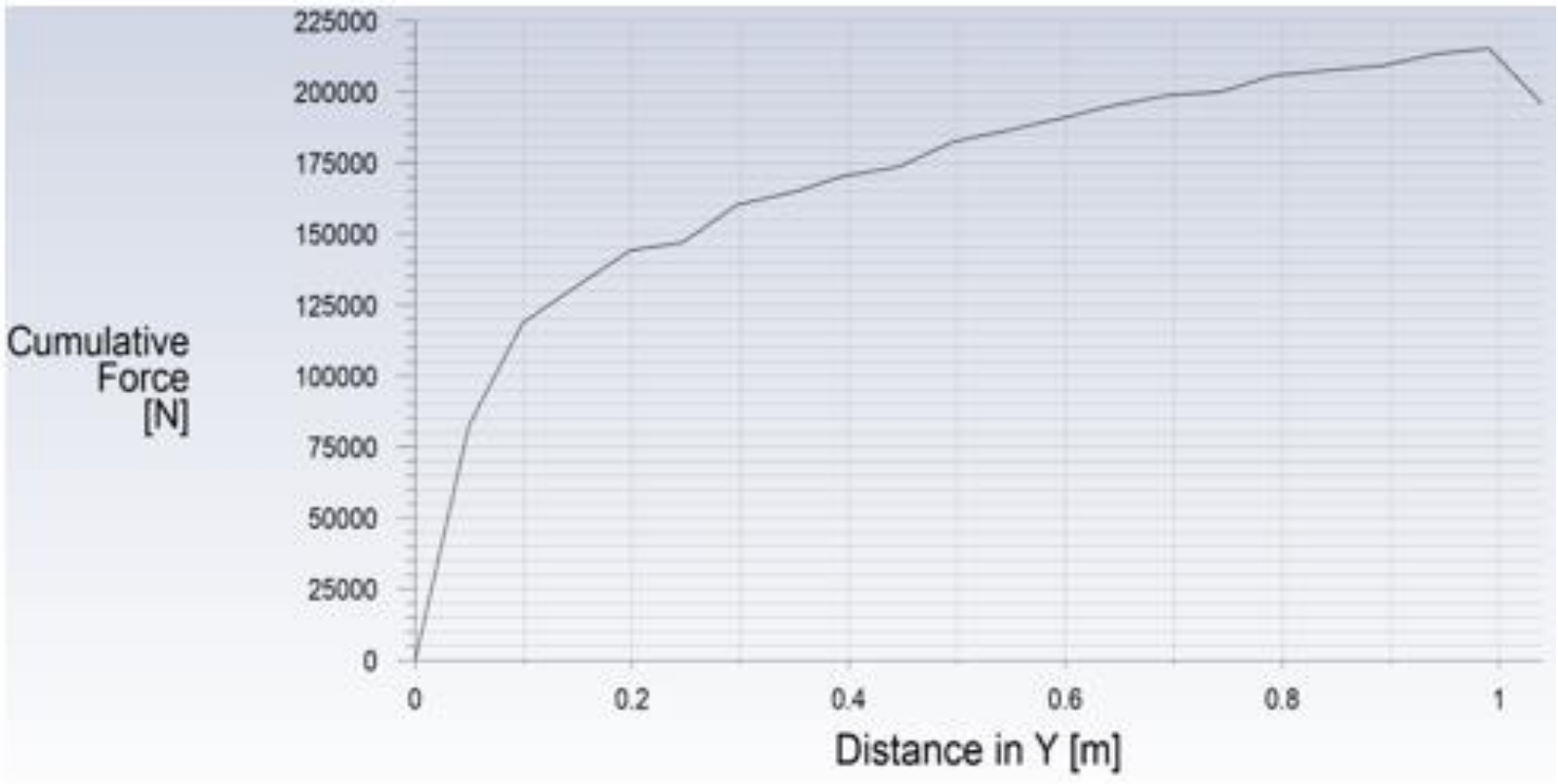

Plot 2: Design.7 drag force by scoop height

An analytical calculation was performed to estimate the force for Design 7, which gets 200 kN compared to a computational value of 195 kN.

Force (by hand) = 2.0e5 N/m    Force (2D model) = 1.95e5 N/m

For Design.8, the top and bottom flat areas will be removed. There will probably be a drag loss as water fails to make direct contact with a perpendicular surface, but how much will it affect fluid flow, vorticity, and velocity. From Design.7, guard rails are a qualitative improvement of over 40% in force.

**Design.8 at 100 m/s**

Design.8 will also have guard rails, but the small flat surfaces at the top and bottom will be smoothed over with the arc of the scoop. In Figure 17, the red circles show the changes from Design.7. The illustration to the far left makes the changes more evident. The average element quality is consistent with Design.6 and Design.7. The lowest element quality is improved without the small flat surfaces.

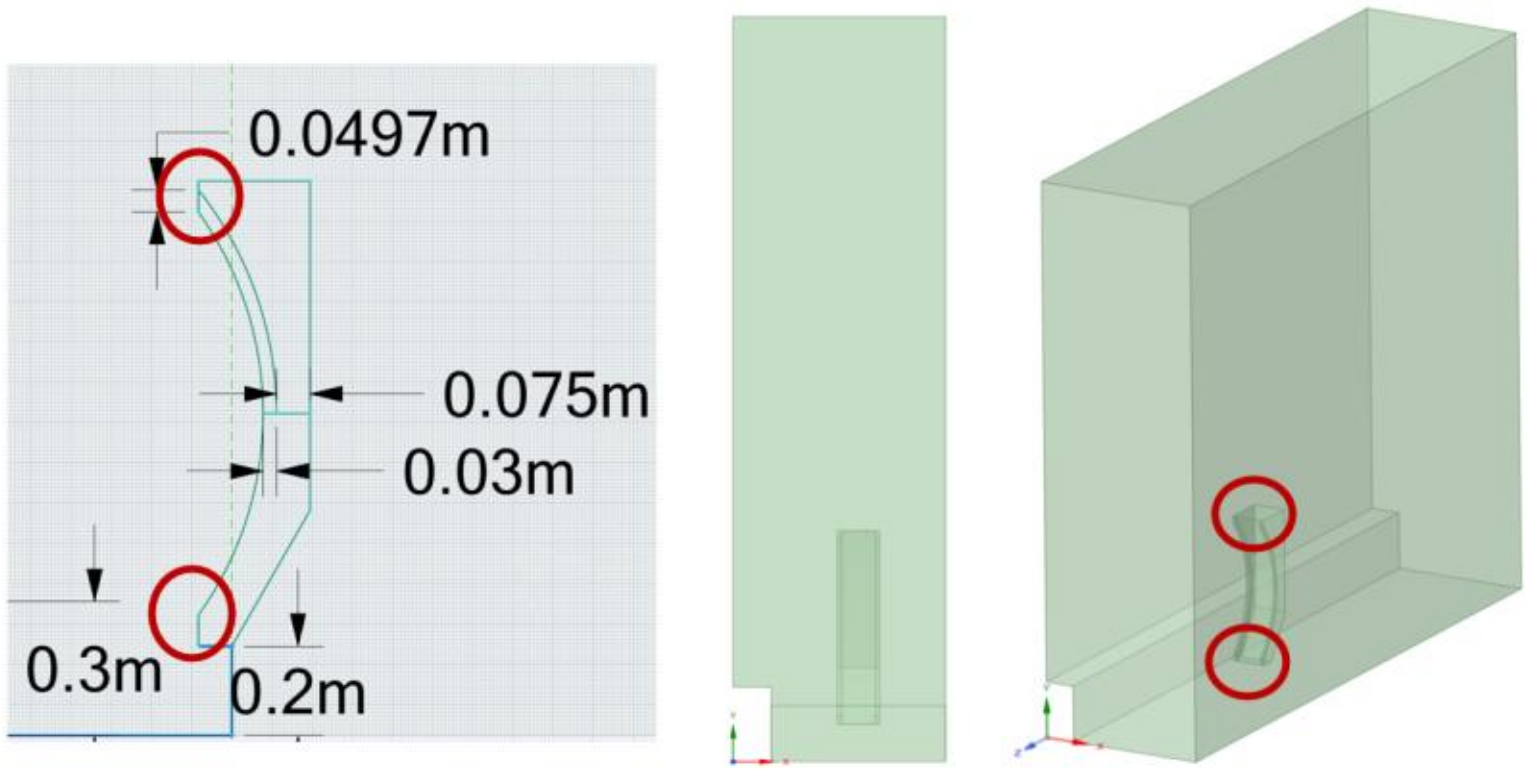

Figure 17: Design.8 geometry changes and views from changes made from Design.7

The changes reduced pressure at the top and bottom of the front scoop by more than 50% (Figure 18), going lower than even Design.6. The volume fraction contour in Figure 19 provides early insight into the flow deflected forward except for the back bottom of the scoop, where vorticity is also at its highest (Figure 19). Before Design.7, vorticity was also at its peak around the front arc and front top.

In conjunction with vorticity, turbulent kinetic energy was noticeably lower (Figure 19), but turbulent intensity behind the scoop was increased (Figure 19). Smooth flow up the scoop has previously removed the turbulence generated from the flat surfaces.

Figure 20 shows the force on the scoop at 179kN, a drop of about 8.3% from Design.7. This drop in energy can be attributed to the channeling of water upwards without perpendicular surfaces disrupting the flow and causing higher turbulence. The lower turbulence may also cause increased flow separation, affecting flow interaction behind the scoop.

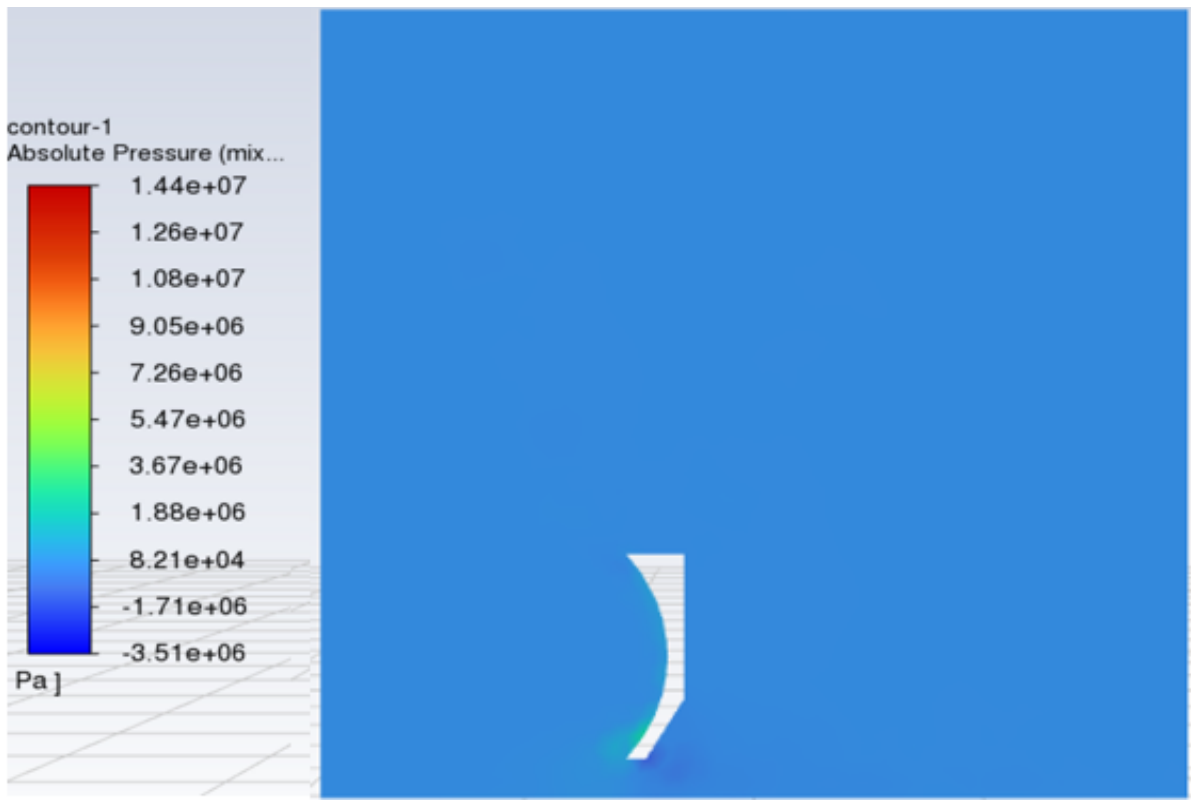

Figure 18: Design.8 absolute pressure contour

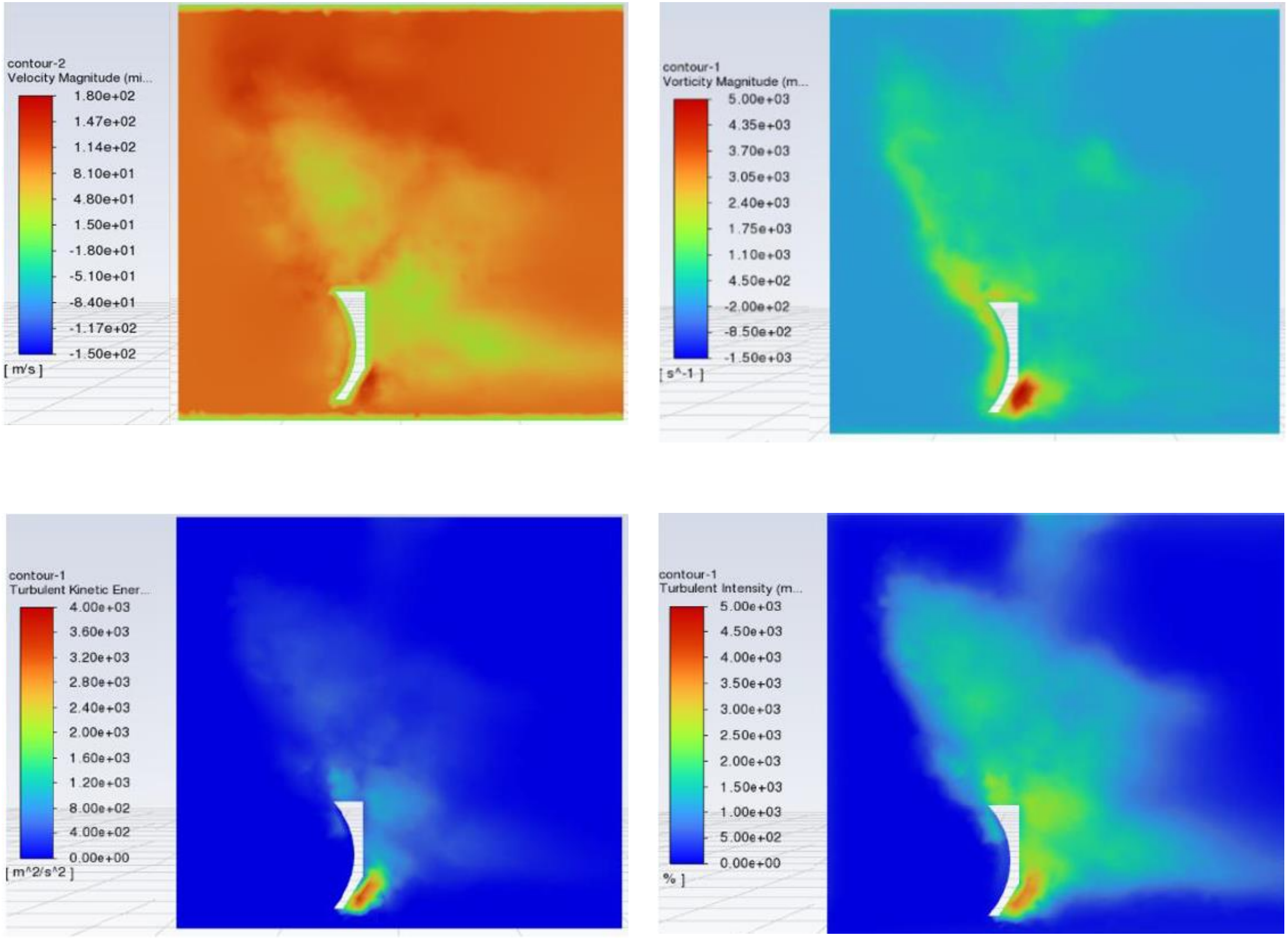

Figure 19: Design.7 (a) velocity magnitude contour (b) vorticity magnitude contour (c) turbulent kinetic energy contour (d) turbulent intensity contour

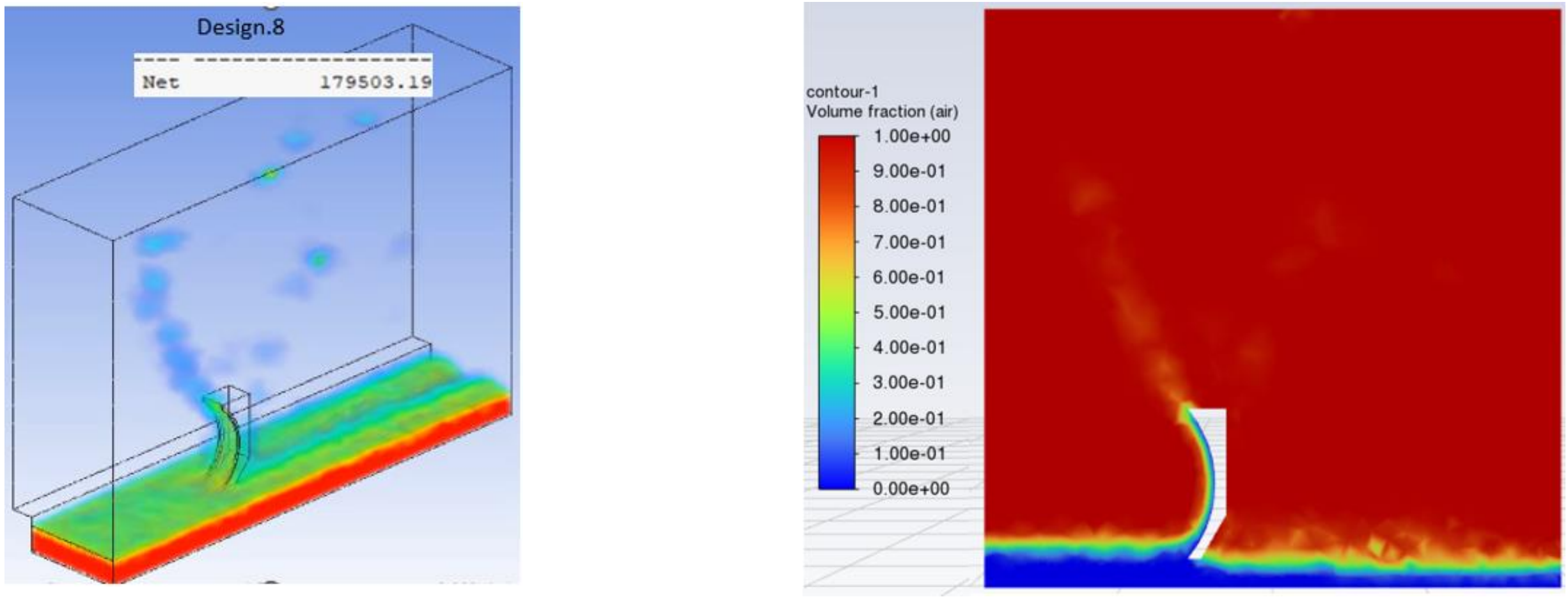

Figure 20: Design.8 (a) water volumetric flow rate (isometric view) (b) volume fraction of air in a plane through the middle of the domain

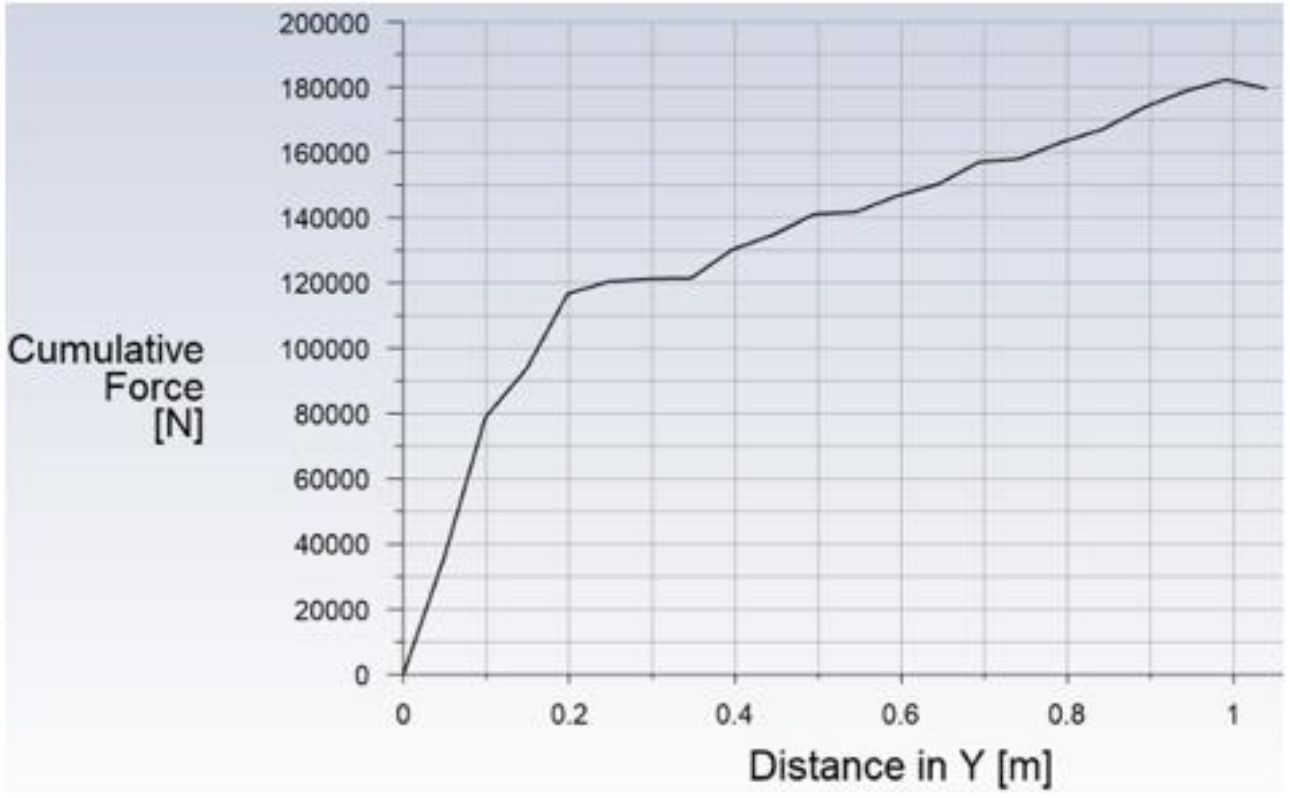

Plot 3: Design.8 drag force by scoop height

Aggregating the results for Design.8, the scoop has less drag and turbulence with sharper increases in velocity vectors on the Z and Y axes. Analyzing Plot 3, the force decline observed in Plot 2 for Design.7 is not as marked. Although Design.7 experiences more drag, the following geometric change will focus on the back top of the scoop.

**Design.9 at 100 m/s**

For Design.9, the top of the back of the scoop will have a 60-degree divergence, as shown in Figure 21. The objective is to analyze how the geometric angled opening affects pressure gradients and flow behind the scoop. Additional benefits include less mass as a scoop and, therefore, weight, which means the scoop and sled could accelerate faster.

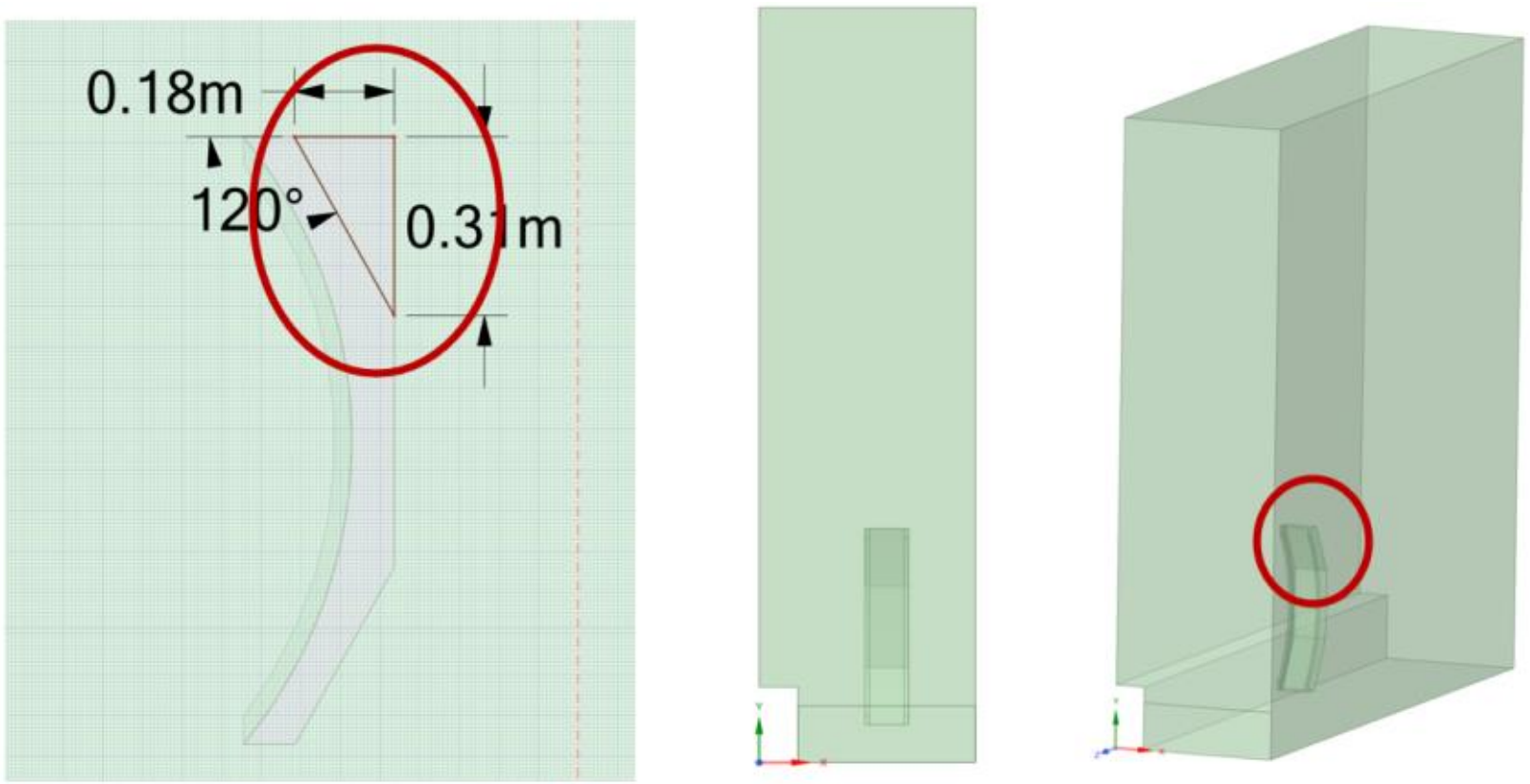

Figure 21: Design.9 geometry changes and views from changes made from Design.8

With the first contour and plot, Figure 22, evident changes in gradient pressure fields are qualified and quantified. The negative pressure point behind the scoop is less, comparatively. This is likely due to less flow separation caused by this change.

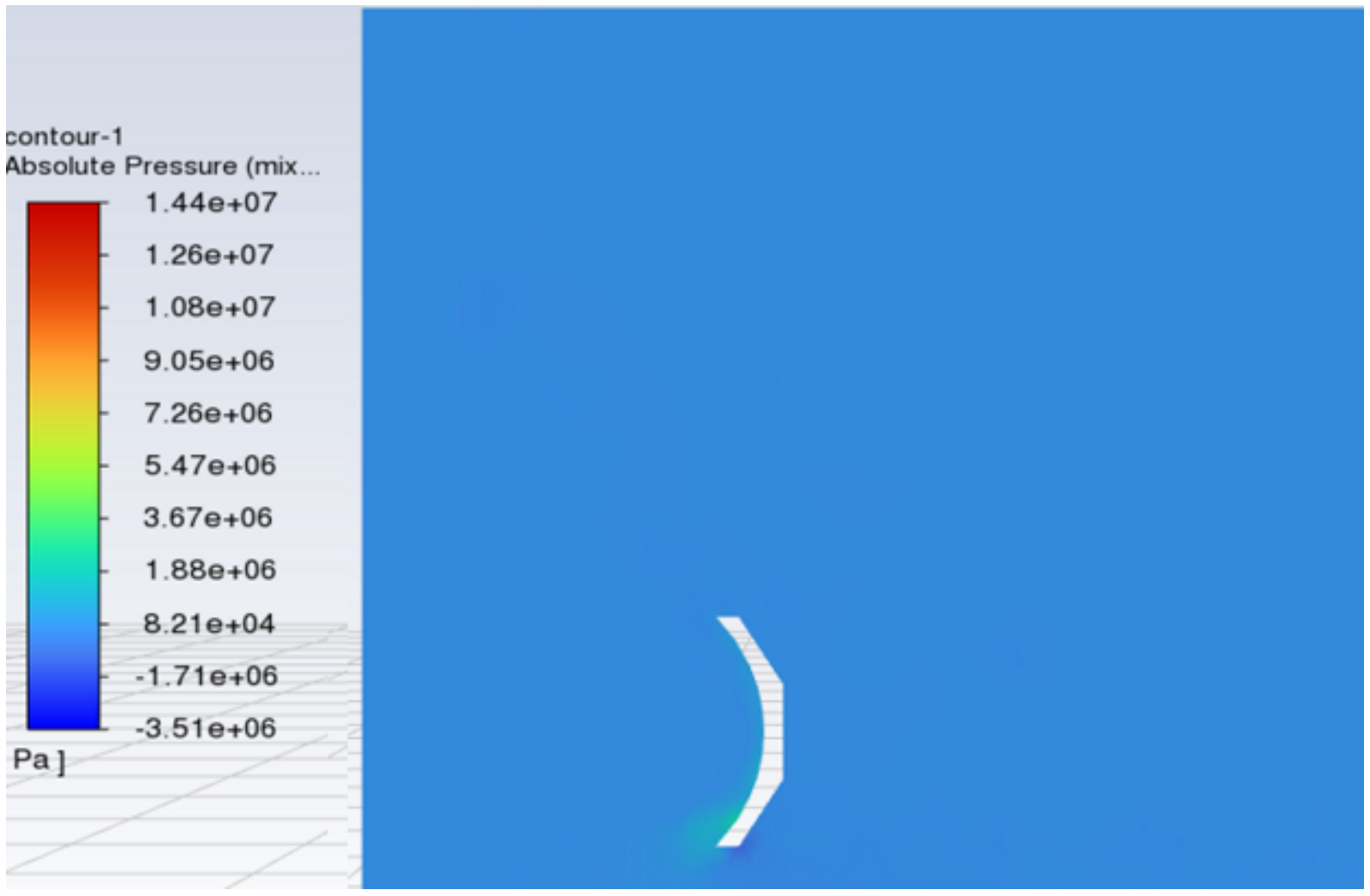

Figure 22: Design.9 absolute pressure contour

Velocity distribution as a magnitude is also more evenly distributed behind the scoop; there are no longer as clear spikes in velocity points as in Design.8. Vorticity, much like velocity, has also been reduced behind the scoop, Figure 23b. Previously, the highest points of vorticity by a noticeable margin were behind the bottom of the scoop. In resemblance with velocity and vorticity, turbulent kinetic energy seems tamer, Figure 23c, although still with a focus on the bottom of the scoop, which is expected. As shown in Figure 23d, the unsteadiness sharply declines compared to previous designs with a flat top scoop across. Turbulent intensity sharply declines from about the 1.7-meter position to the 2-meter position, not seen in the others.

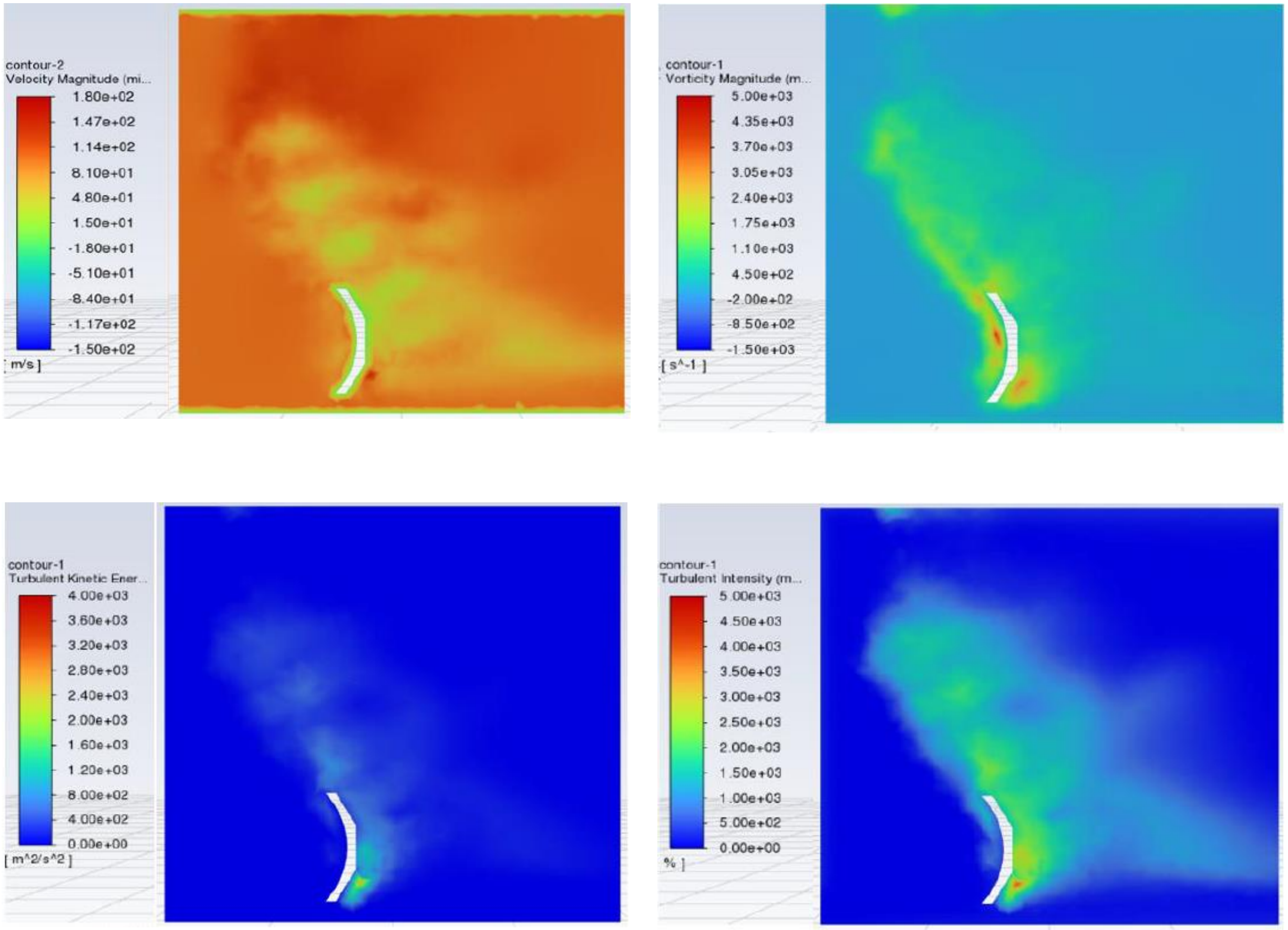

Figure 23: Design.9 (a) velocity magnitude contour (b) vorticity magnitude contour (c) turbulent kinetic energy contour (d) turbulent intensity contour

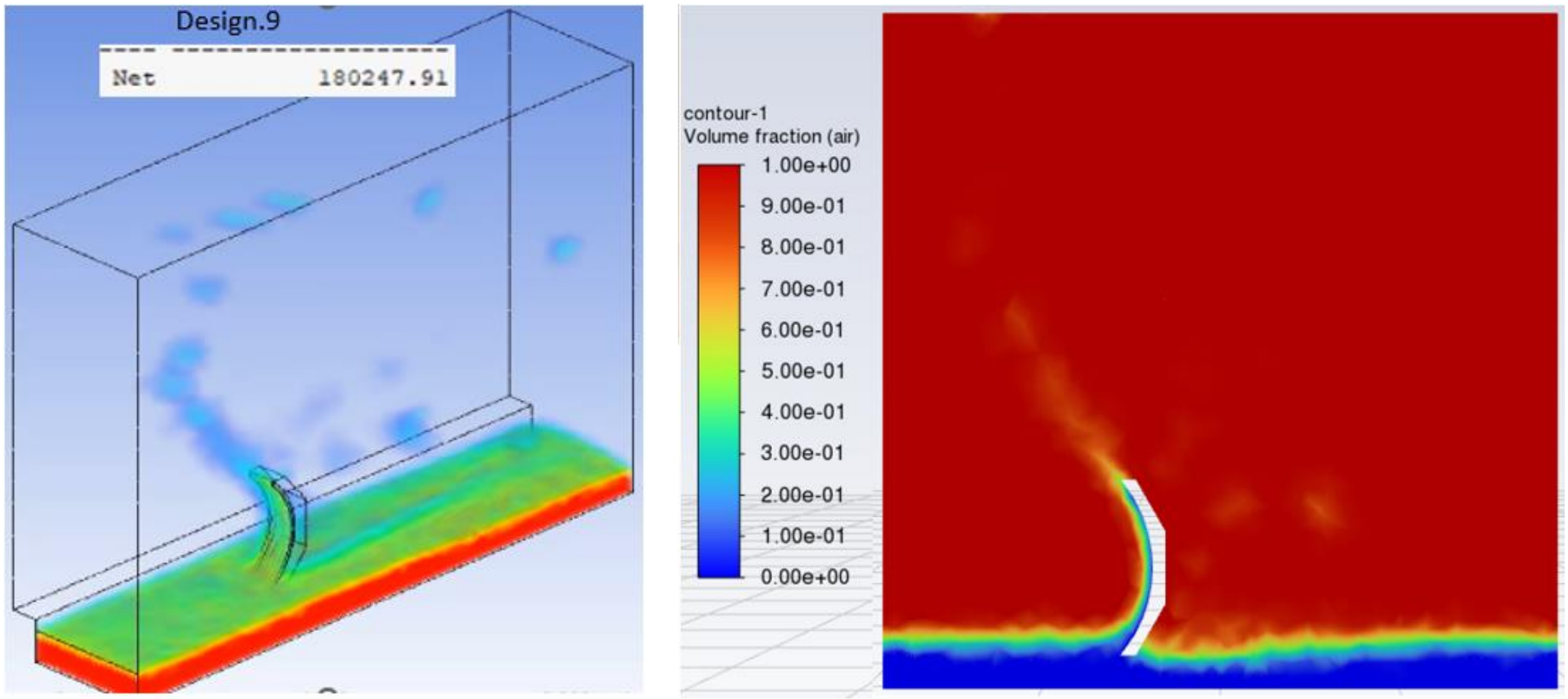

Figure 24: Design.9 (a) water volumetric flow rate (isometric view) (b) Volume fraction of air in a plane through the middle of the domain

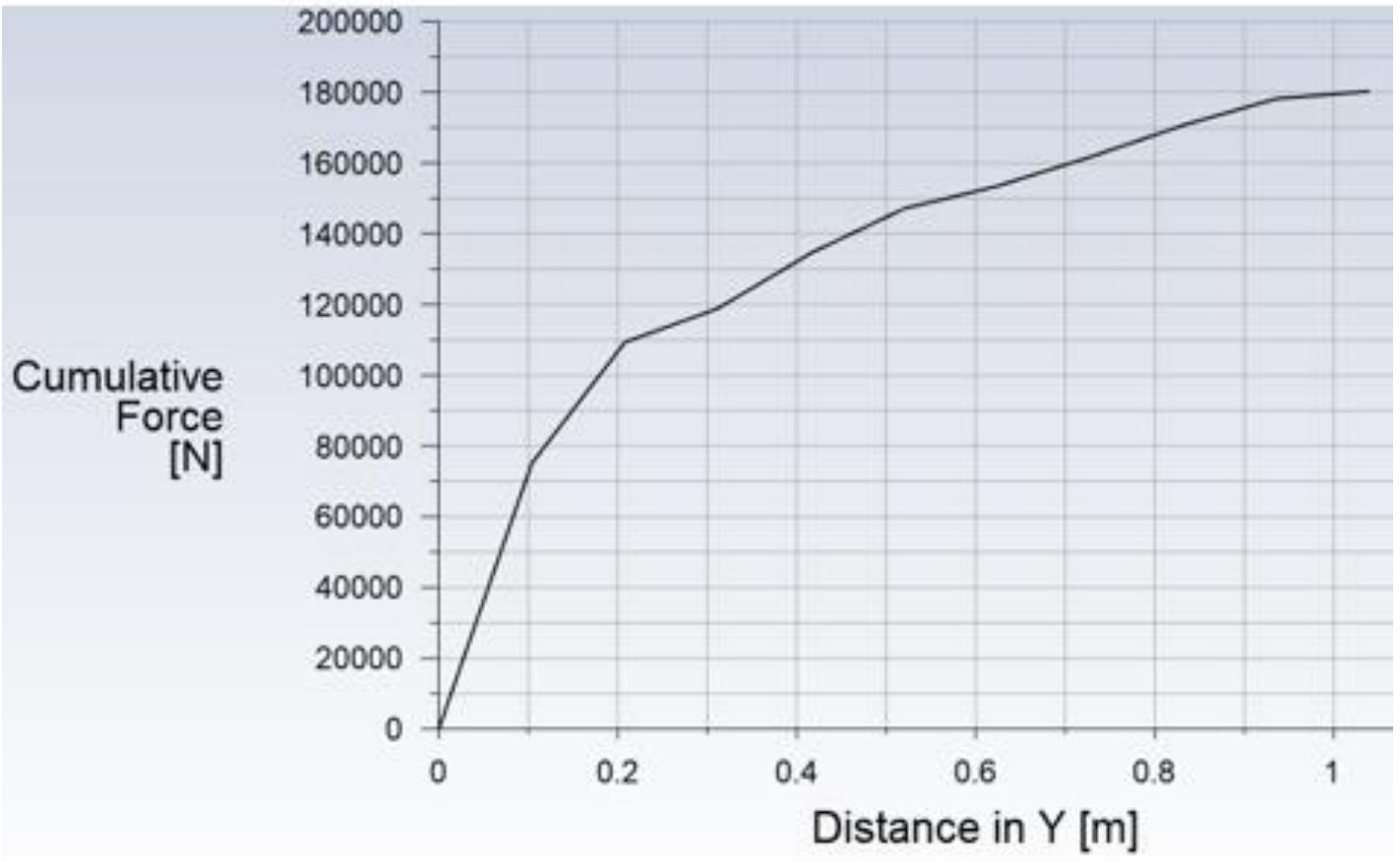

Plot 4: Design.9 drag force by scoop height

The force on the scoop was about 180kN, the same as in Design. 8. The marked difference is the flow behind and past the scoop. This will be essential as the pusher sled is wholly integrated with the scoop. Looking at the water volumetric flow rate in Figure 24, the water flow is substantially different from that of the Design. 7 and Design. 8.

**Design.10 at 100 m/s**

For Design.10, the flow at the bottom front of the scoop is studied by cutting it into a “v”. Additionally, the top back part of the scoop is once again flat as in Design.8. These changes are for two reasons.

1) Though it is known that the drag force will be reduced due to a smaller contact area with the water, quantifying it and knowing how much is important in designing an optimal scoop in which tradeoffs will be made.
2) The scoop is just part of the pusher sled; it is important to observe and understand how it will affect the fluid flow behind it and consider water braking in the context of the entire sled.

The top back part of the scoop seems to cause a drop of force on the scoop; further testing will validate the hypothesis.

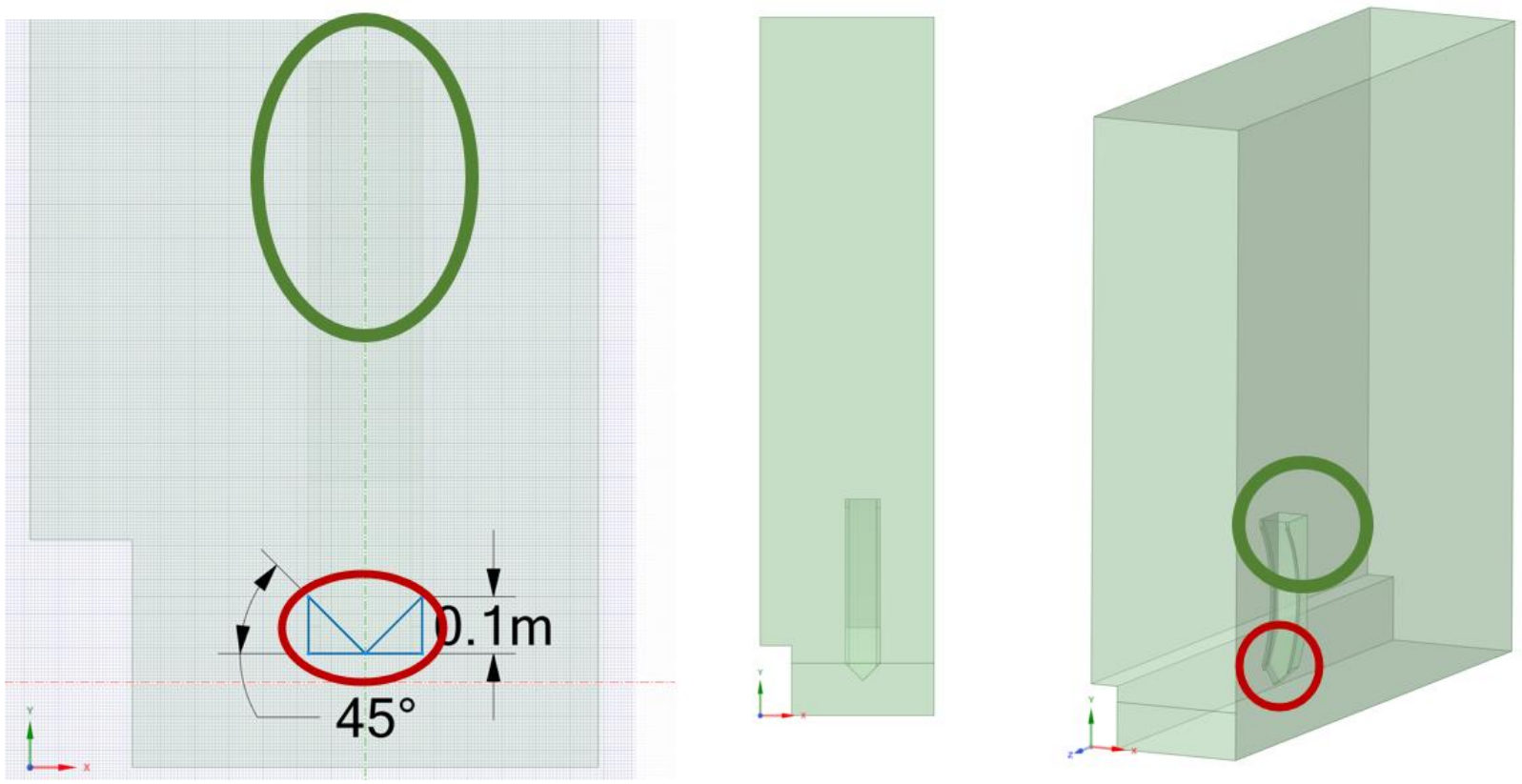

Figure 25: Design10 geometry changes and views from changes made from Design.9

As seen in Figure 25, the “v” cut in the front bottom of the scoop will be at 45-degree angles. An initial look at pressure data, Figure 26, confirms the expected drop from the geometric change of reducing the contact area between the scoop and water. Every metric from Figure 27a.b.c shows an apparent qualitative and quantitative reduction. The only exception is turbulent intensity in Figure 27d. Design.10 from the data has more unsteadiness behind the scoop.

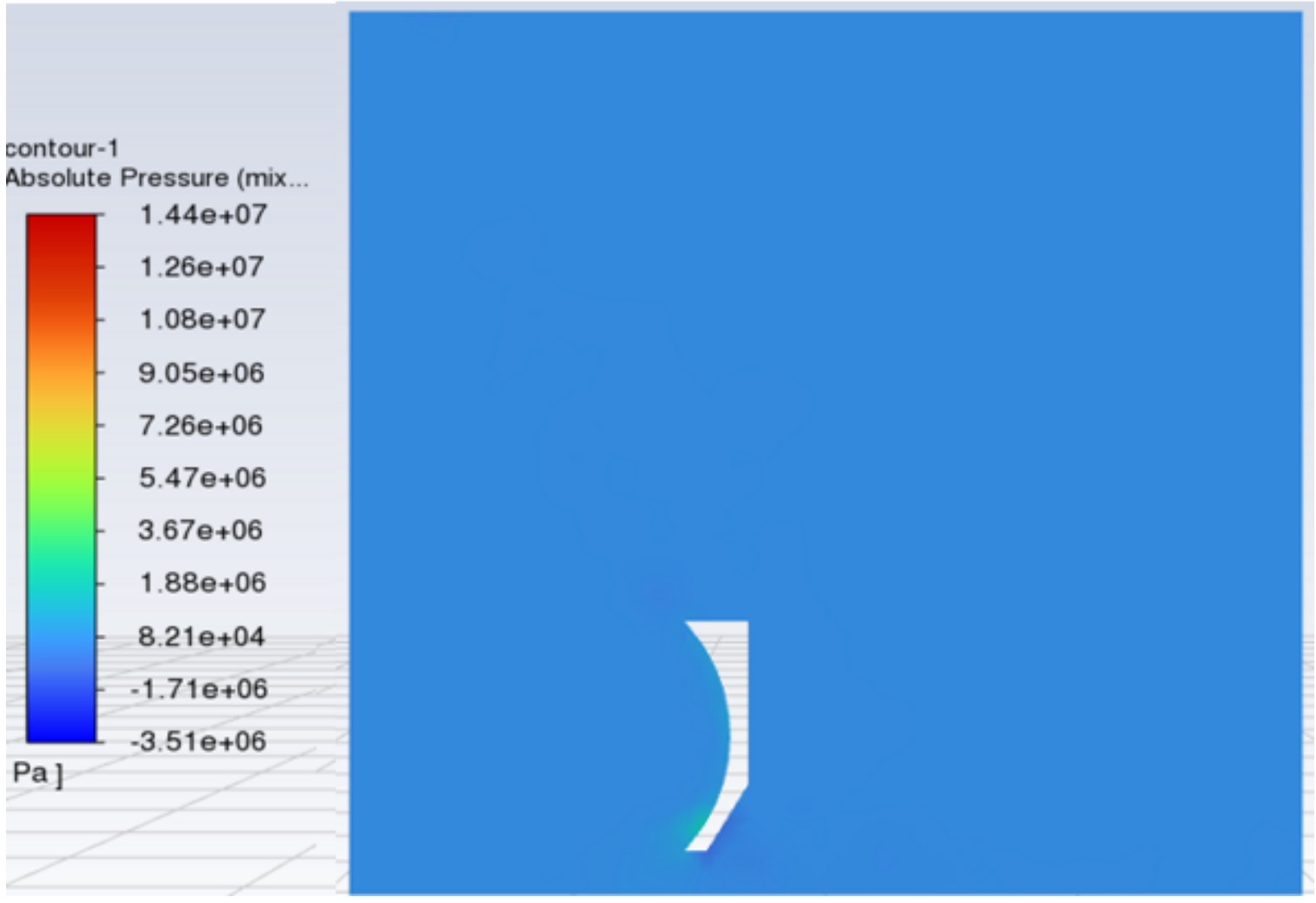

Figure 26: Design.10 absolute pressure contour

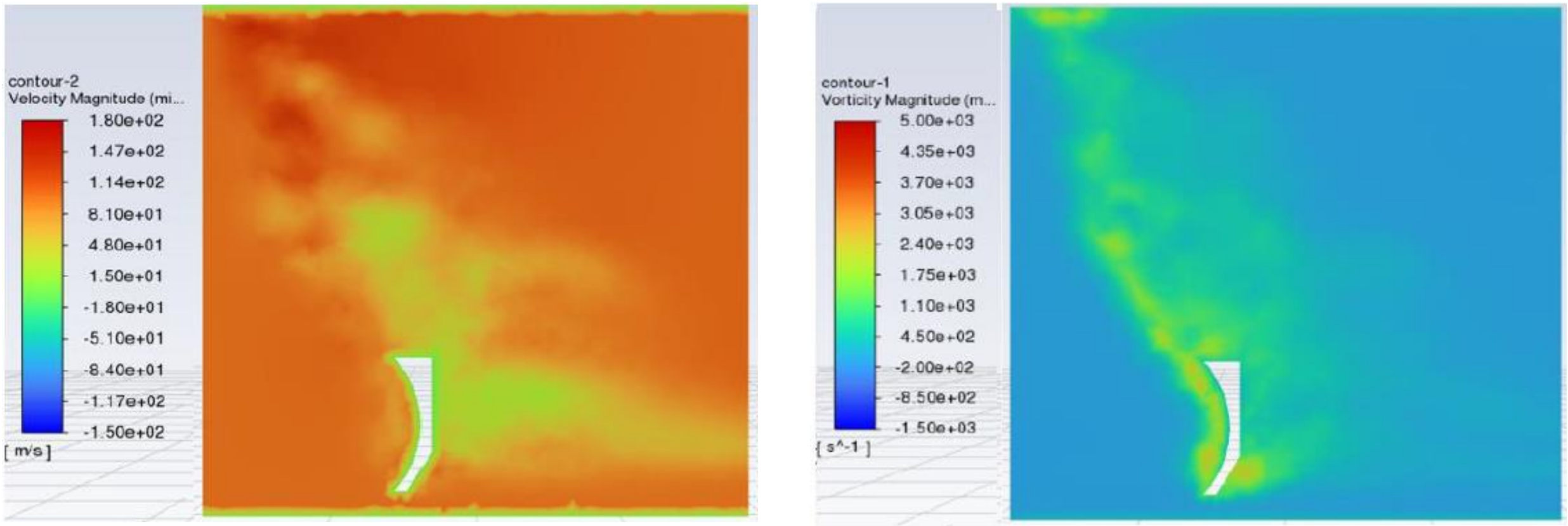

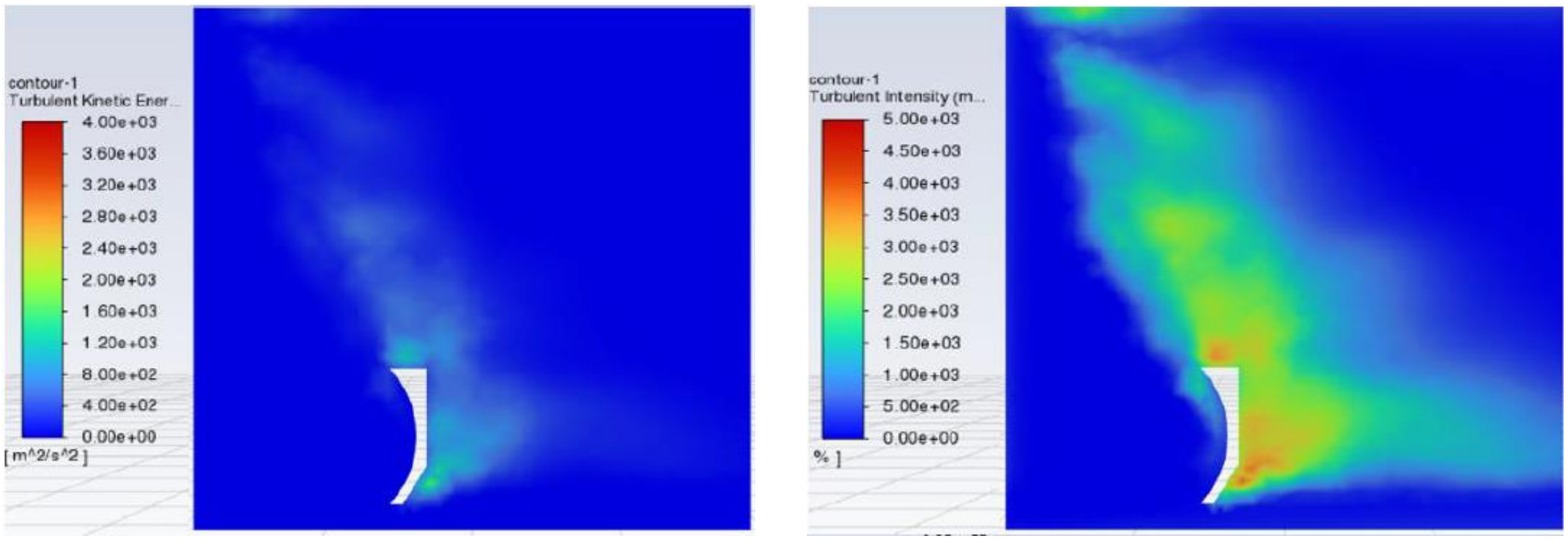

Figure 27: Design.10 (a) velocity magnitude contour (b) vorticity magnitude contour (c) turbulent kinetic energy contour (d) turbulent intensity contour

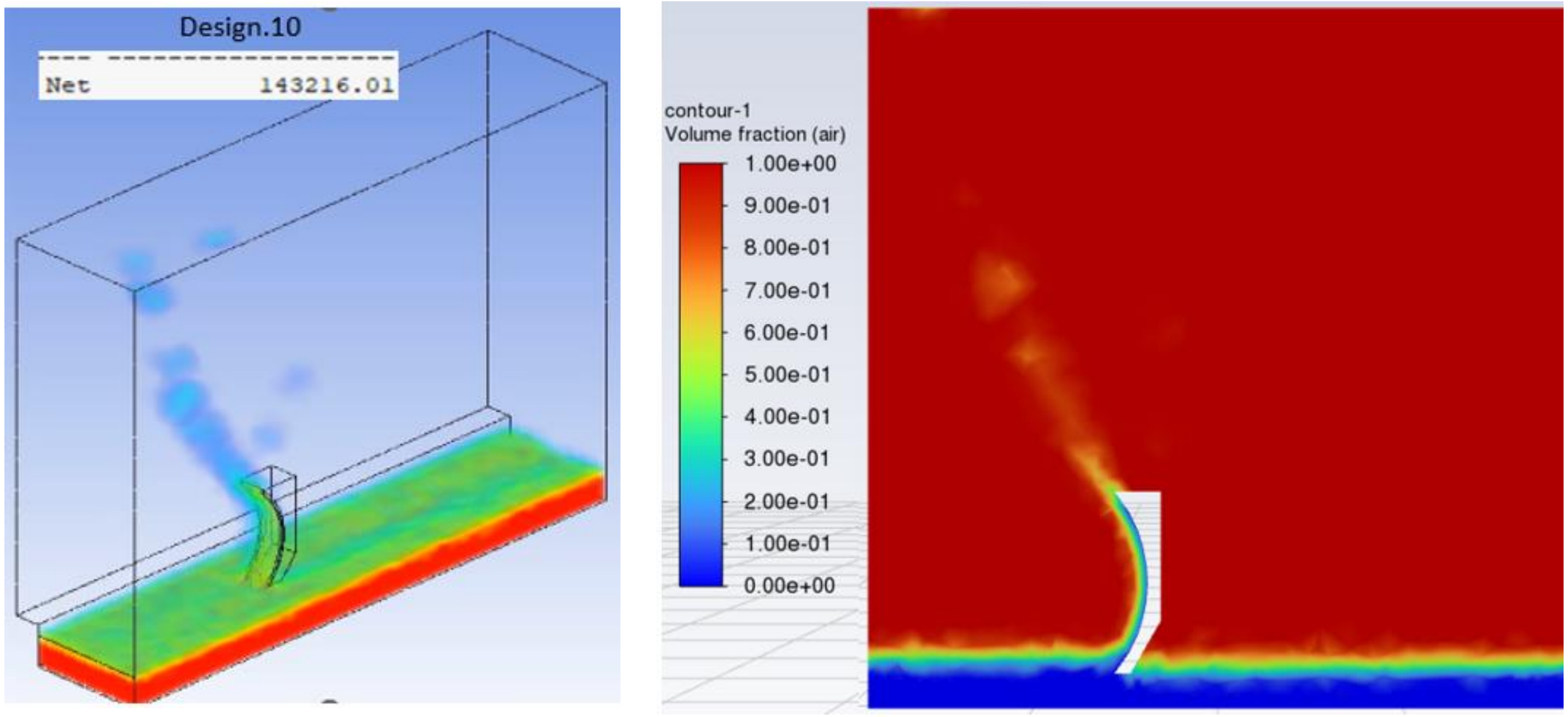

Figure 28: Design.10 (a) water volumetric flow rate (isometric view) (b) Volume fraction of air in a plane through the middle of the domain

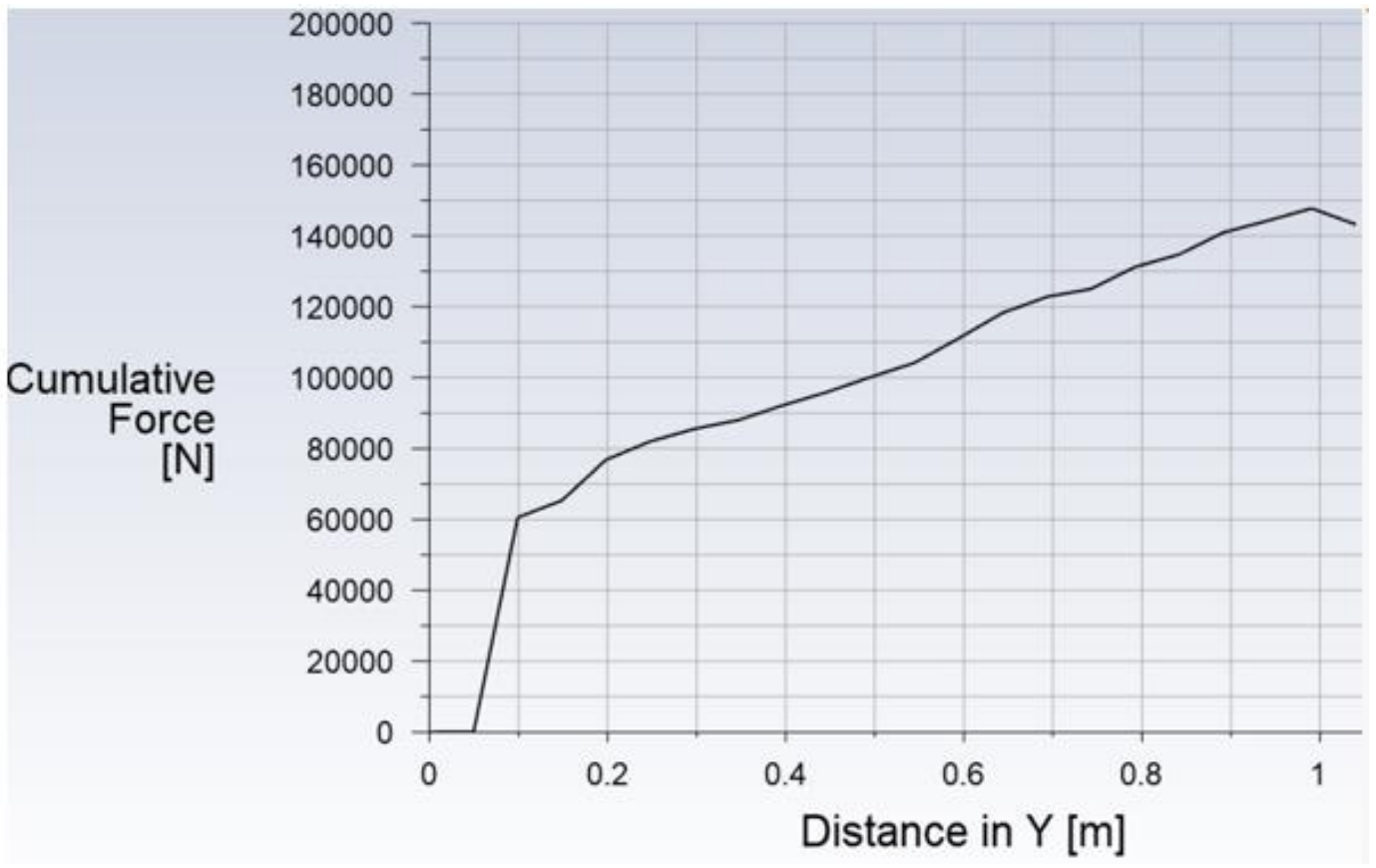

Plot 5: Design.10 drag force by scoop height

Force on the scoop was about 143kN, so about the same as Design.6. This demonstrates that the guard rails are a significant boost to water drag as water contacting a little more than half the area in Design.10 as in Design.6 has about 2.8% more force, 143kN to 139kN. Hand calculation validation at the 0.1-meter height of the scoop produces about 50kN compared to computational 60kN. The significant difference in the calculation is the area estimate in the hand calculation. A drop from 1 meter to 1.04 meters in Plot 5 suggests that a flat top reduces the drag force acting on the scoop.

Careful consideration from analyzing the entire dataset available for Design.10 suggests little benefit in adopting the “v” geometric design for increased drag. It reduces drag considerably from all designs apart from Design.6, and fluid flow behind the scoop will not generate any additional drag from water flow interaction behind the scoop. The only reason for consideration would be to address issues beyond drag, such as lift and less chaotic flow that may damage the pusher sled attached to the scoop.

Next is an analysis of Design.9, which did not fully converge at 300 m/s but did enough to glimpse fluid flow near Mach speed. The final design recommendations and design uses will be made in the Conclusion section.

**CONCLUSION**

All geometric differences offer advantages, depending on what is desired. Design.7 introduces side guards that channel flow very effectively, increasing drag by 40%. In Designs 8, 9, and 10, removing the small flat perpendicular surface from the flow reduced drag by about 8%. The cut in the bottom front of the scoop in Design.10 further reduced force by about 20%. Depending on the objective, recommendations will be presented.

Due to the pusher sled's high reusability requirement, the design must be robust enough to perform under various testing velocities with different attached weights. It must also provide braking capability at high speeds and be durable enough to last many tests. Below are four considerations when optimizing scoop and sled design.

1) Drag force braking sled and bringing it to a stop within a limited amount of distance
2) Lift force on the sled
3) Roll force on the sled
4) Damage to scoop and sled requiring maintenance and replacement.

This study aims to simulate different scoop features that affect fluid flow and make optimal scoop recommendations based on knowledge gained from the various designs. CFD can provide major design insight for the first three on the list above. Number four is limited to fluid flow and would require a fluid-structure interaction (FSI) study.

The measures of pressure, force, velocity, vorticity, turbulent kinetic energy, and turbulent intensity acquired through each design were geared toward designing the optimal geometry of a scoop. Pressure is correlated with force and velocity in magnitude, and all axes offer information about fluid flow behavior. Vorticity in magnitude and all axes provides information on eddy rotations and is correlated to turbulence, which will impact forces in all directions and fluid flow. Data on velocity and vorticity in all axes helps gather perceptions of lift and roll caused by the scoop. Turbulent kinetic energy describes flow intensity, and turbulent intensity measures the unsteadiness of the flow. They are all necessary for design consideration as they directly impact the list of four.

If the objective is to be conservative with the design and optimize for longevity and reduced risk of sled coming off the track, Design.10 would be the best with the additional feature of Design.9 60-degree angle trim from the back top part of the scoop. All quantified measures would be smaller and more stable. There would be a much smaller chance of lift, roll, and damage to the sled, but it would come at the expense of the scoop's primary function, which is to provide drag to stop the sled, about 20% less. The 20% deficit could be recovered by adding Design.7 perpendicular surfaces at the top and bottom, but this would cause more turbulence. That would be the safest optimal design.

For maximum drag, I would recommend Design.9 with Design.7 perpendicular surfaces, increasing drag for Design.9 by about 8%. To offset the increase in turbulence that can cause lift, roll, and damage, the Design.9 back top cut of 60-degree feature works best. To prove that Design.9 reduces turbulence and suffers no decline in force from Design.7 and Design.8 flat back top. Another advantage of Design.9 is that the drag produced would be the most out of all the designs once integrated with the pusher sled. The reason for this is the volumetric flow rate of the water contour. Design 8 and 9 volume fraction figures show that a larger volume of water would contact the pusher component of the sled and cause additional drag. It would come at the expense of potential damage and maintenance requirements, and an FSI study would be required.

**ACKNOWLEDGMENT**

We acknowledge the Holloman High-Speed Test Track team's guidance in this work.

**FUNDING**

The National Science Foundation Graduate Research Fellowship Program (NSF GRFP). The Air Force Office of Scientific Research funded this research under the Agile Science of Test and Evaluation (T&E) program. The U.S. Department of Energy Minority Serving Institutions Partnership Program (DOE-MSIPP) funded this research under grant number GRANT13584020.